\documentclass[prl,aps,twocolumn,superscriptaddress,notitlepage,10pt
]{revtex4-2}
\usepackage{babel}
\usepackage[T1]{fontenc}
\usepackage{amssymb,amsfonts,amsmath,amstext,graphicx}
\usepackage{amsmath}
\usepackage{tikz}
\usepackage{svg}
\usepackage[normalem]{ulem}
\usepackage{tabularx}
\def\bra#1{\langle{#1}|}
\def\ket#1{|{#1}\rangle}
\def\braket#1{\langle{#1}\rangle}

\def\BraVert{\egroup\,\mid\,\bgroup}

\usepackage{ulem}
\usepackage{braket}
\usepackage{xcolor}
\usepackage{ marvosym }
\usepackage[colorlinks, linkcolor=red,urlcolor=blue,citecolor=blue]{hyperref}
\usepackage{orcidlink}
\usepackage{algorithm}
\usepackage{algpseudocode}

\usepackage{graphicx}

\begin{document}
\title{Phase transitions in first-detection statistics of monitored long-range quantum walks}

\author{Sayan Roy \orcidlink{0009-0005-5805-3791}} 
\affiliation{Theoretische Physik, Universit\"{a}t des Saarlandes, D-66123 Saarbr\"{u}cken, Germany} 
\author{Shamik Gupta \orcidlink{0000-0002-6080-4890}}
\affiliation{Department of Theoretical Physics, Tata Institute of Fundamental Research, Homi Bhabha Road, Mumbai 400005, India}
\author{Giovanna Morigi  \orcidlink{0000-0002-1946-3684}}
\affiliation{Theoretische Physik, Universit\"{a}t des Saarlandes, D-66123 Saarbr\"{u}cken, Germany}
\affiliation{Center for Quantum Technologies (QuTe), Saarland University, Campus, 66123 Saarbr\"ucken, Germany}
\author{Gabriele Perfetto \orcidlink{0000-0002-4568-2311}}
\affiliation{Institut für Theoretische Physik, ETH Zürich, Wolfgang-Pauli-Str. 27, 8093 Zürich, Switzerland}

\begin{abstract}  
In a quantum walk, the first-detection return probability (FDRP) characterizes salient features, determining whether the quantum walk is transient or recurrent. We study the FDRP of quantum walks on a chain where the initial site is stroboscopically monitored by a detector and the walker performs long-range hopping between sites. We assume that the hopping strength decays with the distance $d$ as $d^{-\alpha}$  and $\alpha\geq 0$ and show that the power-law exponent $\alpha$ critically determines the behavior of the FDRP. The value $\alpha=1$ separates recurrent ($\alpha<1$) from transient ($\alpha>1$) quantum walks  through a continuous phase transition in the total detection probability. For $\alpha<1$,  strong long-range hopping induces localization, resulting in unit total detection probability. Instead, for $\alpha>1$ the long-range walk is transient and the return probability decays algebraically as a function of time as $t^{-\beta}$. The associated decay exponent $\beta$ features nonanalytic points as a function of $\alpha$. Such singularities are not exclusively determined by the low-energy spectrum, but are caused by the interference between infrared and ultraviolet energy modes induced by projective measurements, signalling the emergence of critical behavior intrinsic to the non-unitary dynamics. These dynamics are solely controlled by tuning the long-range exponent $\alpha$ and can thus be experimentally probed in atomic and molecular systems.
\end{abstract}

\maketitle

\emph{Introduction.---} Random walks constitute a fundamental paradigm of physics. Therein, recurrence and transience are key concepts since they quantify the ability of the walker to return to the starting point \cite{chandrasekhar1943}. This ability is described  by the celebrated P\'olya's theorem~\cite{Polya1921}, which states that in one and two dimensions, the walk is \emph{recurrent}, namely, the walker returns to the initial site with unit probability. In higher dimensions, instead, return is not guaranteed and the walk is said to be \emph{transient}. The distinction between recurrent and transient dynamics is captured by the first-return probability of the walker to its initial site, see, e.g., Refs.~\cite{redner2001guide,bray2013persistence,Sandev_book}. This classification is paradigmatic and has far-reaching consequences in classical physics: it relates to the efficiency of search problems \cite{IntermittentSearch,Optimal_search_1},  controls the kinetics of diffusion-limited reactions \cite{toussaint1983particle,hinrichsen2000non,tauber2014critical}, and explains the absence of spontaneous symmetry breaking for continuous symmetries in dimensions less than or equal to two, as given by the Mermin-Wagner theorem~\cite{Cassi1992}.

Recurrence becomes more subtle in the quantum domain because its definition requires the introduction of a detector at the initial site, which necessarily perturbs the walker's subsequent evolution. Projective measurements at stroboscopic times yield a binary outcome, success or failure, corresponding to whether the walker gets detected or not. The ensuing long-time dynamics have been shown to behave in a markedly different manner compared to classical walks~\cite{QuantumWalksTodd2006,grunbaum_recurrence_2013,Dhar2015QuantumModel,dhar2015b,Barkai2016,Friedman2017QuantumProblem,Barkai2018,Barkai2020,Dittle_fdt_many_body,Thiel_spectral,imparato_many_body_fdt,Walter2025ThermodynamicSystems}. Quantum walks on a one-dimensional lattice are transient when the hopping is to nearest neighbors. The associated first-detection return probability (FDRP) $F_n$ decays at long times algebraically with the stroboscopic time $n\tau$ as $F_n \sim n^{-3}$, where the exponent is double of that in the classical walk. This highlights a fundamental contrast with classical dynamics: quantum interference and measurement back-action suppress the probability of late-time detection. The emerging dynamics can be relevant for spatial searches based on quantum walks \cite{quantum_walk_grover,Farhi_Gutmann98,quantum_walk_spatial_search,reset_geometric_brownian,tornow2023,yin2025restart,yin2025resonance,liu2026fractionally,King2025_optimalsearch,king2026timecomplexity}, network dynamics \cite{Barabasi:2016_review, santhanam2026,Folz:2023}, and quantum algorithms \cite{Aaronson:2005, Kendon2010, Childs:2010_review}.

The origin of transience under repeated projective measurements can be traced back to the spectral properties of the unitary evolution~\cite{grunbaum_recurrence_2013}. This can be understood through the so-called quantum renewal equation \cite{QuantumWalksTodd2006,grunbaum_recurrence_2013,Dhar2015QuantumModel,dhar2015b,Barkai2016}, which directly links the FDRP with the return-Loschmidt amplitude of the unitary evolution. The asymptotic long-time behavior of the latter is, in turn, determined by the singularities of measurement spectral density of states~\cite{Thiel_spectral}. Long-range non-local hopping, in a related way, modifies infrared properties of the energy spectrum (see Refs.~\cite{defenu_long-range_2023,defenu_out--equilibrium_2024} for reviews on the subject) disrupting linear information spreading \cite{nl_LR_1,nl_LR_2,nl_LR_3,nl_LR_4,nl_LR_5,nl_LR_6}. 
How the interplay of stroboscopic measurements and long-range hopping shapes the spectral properties, and thus determines the nature of the quantum walk, is the core question of this work.

\begin{figure}[t]
\includegraphics[width=\columnwidth]{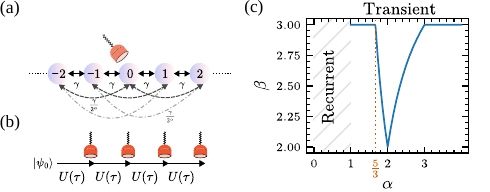}
\caption{(a) A walker coherently hops along a ring with hopping amplitude decaying as $d^{-\alpha}$ with the distance $d$. (b) A detector monitors stroboscopically the initial site $\ket{0}$, a positive detection event terminates the dynamics. We show that the first-detection probability decays algebraically as $F_n \sim n^{-\beta}$, where $\beta$ depends critically on $\alpha$, as visible in (c), identifying different regimes of the transient quantum walk. The nonanalytic points at $\alpha=1,2,3$  reflect a  phase transition in the spectrum of the long-range model. Instead, the nonanalytic point at $\alpha=5/3$ is due to the projective effect of measurements.} 
    \label{fig:1}
\end{figure}

We will show that the behavior of the FDRP of a monitored quantum walk is determined by the interplay of long-range hopping and projective measurements, see Fig.\ \ref{fig:1}(a) and (b). The unitary dynamics is the analogue of a Lévy flight \cite{levy_th_1,levy_th_2,levy_th_3,levy_th_4,levy_th_5}, where the transition amplitude between two sites at distance $d$ decays with $d^{-\alpha}$. The measurement is performed at stroboscopic times. Figure \ref{fig:1}(c) summarizes our insights: the exponent $\alpha\ge 0$ critically determines the nature of the quantum walk, giving rise to a plethora of dynamics and permitting to control the quantum walk properties.

\emph{Monitored long-range quantum walk.---} We consider a single quantum walker constrained to move on a ring of $N$ sites. It is initially localized at one site ($\ket{\psi(0)}=\ket{0}$) and evolves under unitary evolution interspersed by projective measurements at stroboscopic times $t=n\tau$, with $\tau>0$ and $n=1,2,\ldots$, monitoring whether the walker has returned to the initial site, see Fig.~\ref{fig:1}(a) and (b).
The coherent dynamics describes long-range hopping between sites ($\hbar=1$):
\begin{align}
\hat H_\alpha=-\frac{\gamma}{\mathcal N_\alpha}\sum_{i<j}\frac{1}{d_{ij}^{\alpha}}\ket{i}\bra{j}+\mathrm{H.c.}\,,
\label{eq:Hamiltonian_main}
\end{align}
with $\gamma>0$ a scaling factor and $d_{ij}=\min(|i-j|,N-|i-j|)$ the shortest distance on the ring. The exponent $\alpha$ controls the tunnelling range, interpolating between all-to-all ($\alpha\to0$) and nearest-neighbor ($\alpha\to\infty$) hopping. The Kac normalization $\mathcal N_\alpha=\sum_{d=1}^{N/2} d^{-\alpha}$ ensures a well-defined thermodynamic limit $N\to \infty$~\cite{Kastner2011DivergingModels,kastner2025longrangesystemsnonextensivityrescaling}. The Hamiltonian is diagonal in Fourier space with dispersion relation
\begin{equation}
E_{k_l}(\alpha)  =  -  \frac{2\gamma}{\mathcal{N}_\alpha}  \sum_{d = 1}^{\lfloor \frac{N}{2} \rfloor} \frac{1}{d^\alpha} \cos(k_l d),
\label{eq:dispertion_relation_main}
\end{equation}
where $k_l=2\pi l/N$ are quasimomenta ($l=0,1\dots N-1$). See the Supplemental Material(SM)~\cite{SM}.

Due to the measurement, the overall dynamics is not unitary and can be cast into the sum of trajectories conditioned to a detection event \cite{grunbaum_recurrence_2013,Dhar2015QuantumModel,dhar2015b}. The trajectory corresponding to no-click event till time $n\tau$ is given by:
\begin{equation}
\label{eq:theta_n}
\ket{\theta_n} = e^{-i \hat H_\alpha \tau}\left[(\mathbb{I}-\ket{\psi_0}\bra{\psi_0})e^{-i \hat H_\alpha \tau}\right]^{n-1}\ket{\psi_0}.
\end{equation}
Its squared norm is the probability that the walker has not been detected in any of the previous $n-1$ measurements. Correspondingly, the first-detection 
return amplitude is given by $\phi_n=\braket{0|\theta_n}$ \cite{Dhar2015QuantumModel,Friedman2017QuantumProblem} and satisfies the quantum renewal equation
\begin{align}
\phi_n
= \bra{0} e^{-i \hat H_\alpha n \tau}\ket{0}
-\sum_{j=1}^{n-1}\phi_j\bra{0}  e^{-i \hat H_\alpha (n- j) \tau}\ket{0},
\label{eq:renewal}
\end{align}
with the FDRP defined as $F_n=|\phi_n|^2$. The first term is the measurement-free return amplitude also known as Loschmidt amplitude $\mathcal L_n=\bra{0} e^{-i \hat H_\alpha n \tau}\ket{0}$ \cite{Gorin2006,Peres84}. The convolution term contains the negative outcome of the previous $n-1$ measurements. Equation \eqref{eq:renewal} is conveniently solved by introducing the generating functions $\tilde{\phi}(z)\equiv\sum_{n\ge1}z^n\phi_n$ and $\tilde{\mathcal L}(z)\equiv\sum_{n\ge1}z^n\mathcal L_n$ in terms of which one has
\begin{align}
    \tilde{\phi}(z)= \frac{\tilde{\mathcal L}(z)}{1+\tilde{\mathcal L}(z)}.
    \label{eq:phiz}
\end{align}
The first-detection amplitude $\phi_n$ then follows from the inverse $Z$-transform
\begin{align}
\phi_n=\frac{1}{2\pi i}\oint_C dz~\frac{\tilde{\phi}(z)}{z^{n+1}},
\label{eq:inv_z_phi_main}
\end{align}
where $C$ is a contour encircling the origin and lying entirely within the domain of analyticity of $\tilde{\phi}(z)$. Equations \eqref{eq:phiz} and \eqref{eq:inv_z_phi_main} show that the asymptotic behavior of the FDRP is controlled by the nonanalyticities of $\tilde{\mathcal L}(z)$ \cite{Dhar2015QuantumModel,dhar2015b,Barkai2016,Friedman2017QuantumProblem,Walter2025ThermodynamicSystems}. The latter is fully determined by the measurement-free dynamics. 

\emph{Loschmidt amplitude.---} The Loschmidt amplitude $\mathcal L_n$
is analyzed separately in the strong ($\alpha<1$) and in the weak ($\alpha>1$) long-range regime. In the strong long-range regime, the Kac normalization $\mathcal{N}_{\alpha}$ diverges in the thermodynamic limit $N\to \infty$. The resulting energy spectrum \eqref{eq:dispertion_relation_main} is finite and discrete \cite{Defenu2021MetastabilitySystems,defenu_out--equilibrium_2024,SM}. The long-wavelength modes $l\sim \mathcal{O}(1)$ form a sub-extensive set, whose contribution to $\mathcal{L}_n$ vanishes as $N \to \infty$. The short-wavelength modes then dominate, yielding 
\begin{equation}
\mathcal{L}_n \to 1 \quad \mbox{as} \quad N\to\infty \quad \mbox{for} \quad \alpha \leq 1.
\label{eq:strong_LR_loschmidt}
\end{equation}
for any fixed $n$ and $\tau$. Instead, in the weak-long range regime $\alpha>1$, $\mathcal{N}_{\alpha}$ attains a finite limit as $N\to \infty$ and the spectrum becomes continuous. One then has
\begin{align}
\label{eq:ln_alphag1_main}
    \mathcal{L}_n =  \frac{1}{2 \pi} \int_{0}^{2 \pi} d k\,  e^{-i E_k(\alpha) n \tau} \quad \mbox{as} \quad N\to\infty \quad \mbox{for} \quad \alpha>1,
\end{align}
where $E_k(\alpha) = -2 \gamma \rm{Re} \left[Li_\alpha(e^{ik})\right]/\zeta(\alpha)$ from \eqref{eq:dispertion_relation_main} and $\rm Li_\alpha(z)$ denotes the polylogarithm of order $\alpha$ \cite{NIST:DLMF}. For $n \tau\gg 1$, the integrand oscillates fast as a function of $k$. Asymptotic contributions are at regular stationary points $k=k^*$ satisfying $\mathrm{d}E_k / \mathrm{d}k\vert_{k=k^*} = 0$ and at nonanalytic points where the stationary-phase approximation does not directly apply. The ultraviolet band edge at  $k^*=\pi$ is a regular quadratic stationary point for $\alpha>1$ giving rise to the contribution $\mathcal L_n^{(\pi)}\sim (n\tau)^{-1/2}$. The infrared mode at $k^*=0$, instead, is qualitatively modified by the branch-point singularity of the polylogarithm at $e^{i k} = 1$. For $\alpha>3$, it is a regular quadratic stationary point; for $2<\alpha<3$, it is stationary but not quadratic since $\mathrm{d}^2 E_k/\mathrm{d} k^2$ diverges; finally,  for $1<\alpha<2$, $\mathrm{d} E_k/\mathrm{d} k$ diverges, which determines a divergent group velocity and the violation of the Lieb-Robinson bound \cite{nl_LR_1,nl_LR_2,nl_LR_3,nl_LR_4,nl_LR_5,nl_LR_6,SM}. This term gives the contribution $\mathcal L_n^{(0)}\sim (n\tau)^{-f(\alpha)}$ with $f(\alpha) = 1/(\alpha -1)$, so that for $1<\alpha<3$ the asymptotics of the return amplitude reads
\begin{eqnarray}
\mathcal{L}_n\approx  a_0 e^{-in \tau E_0}\,n^{-f(\alpha)} +  a_{\pi}  e^{-in\tau E_\pi} \,n^{-1/2}\,,
\label{eq:Loschmidt_asymp_main_text}
\end{eqnarray}
with $a_0$ and $a_\pi$ being the pre-factors (Eq.~\eqref{eq:Loschmidt_asymp_main_text} contains logarithmic corrections at $\alpha=3$~\cite{SM, Defenu2019}). Consequently, the generating function $\mathcal{L}(z)$ is asymptotically expressed in terms of polylogarithms as
\begin{align} 
\label{eq:lz_polylog_1_alpha_3_main}
    \tilde{\mathcal{L}}(z) \approx a_0 \, \mathrm{Li}_{f(\alpha)}(ze^{-i E_0 \tau}) + a_\pi \, \mathrm{Li}_{\frac{1}{2}}(ze^{-i  E_\pi \tau})\,,
\end{align}
with branch-cut singularities at $z_0=e^{iE_0\tau}$ and $z_\pi=e^{iE_\pi\tau}$. These nonanalyticities control the corresponding branch-cut structure of $\tilde{\phi}(z)$ and therefore the long-time behavior of $\phi_n$ in Eq.~\eqref{eq:inv_z_phi_main}. 
\begin{figure}[!t]
\includegraphics[width=\columnwidth]{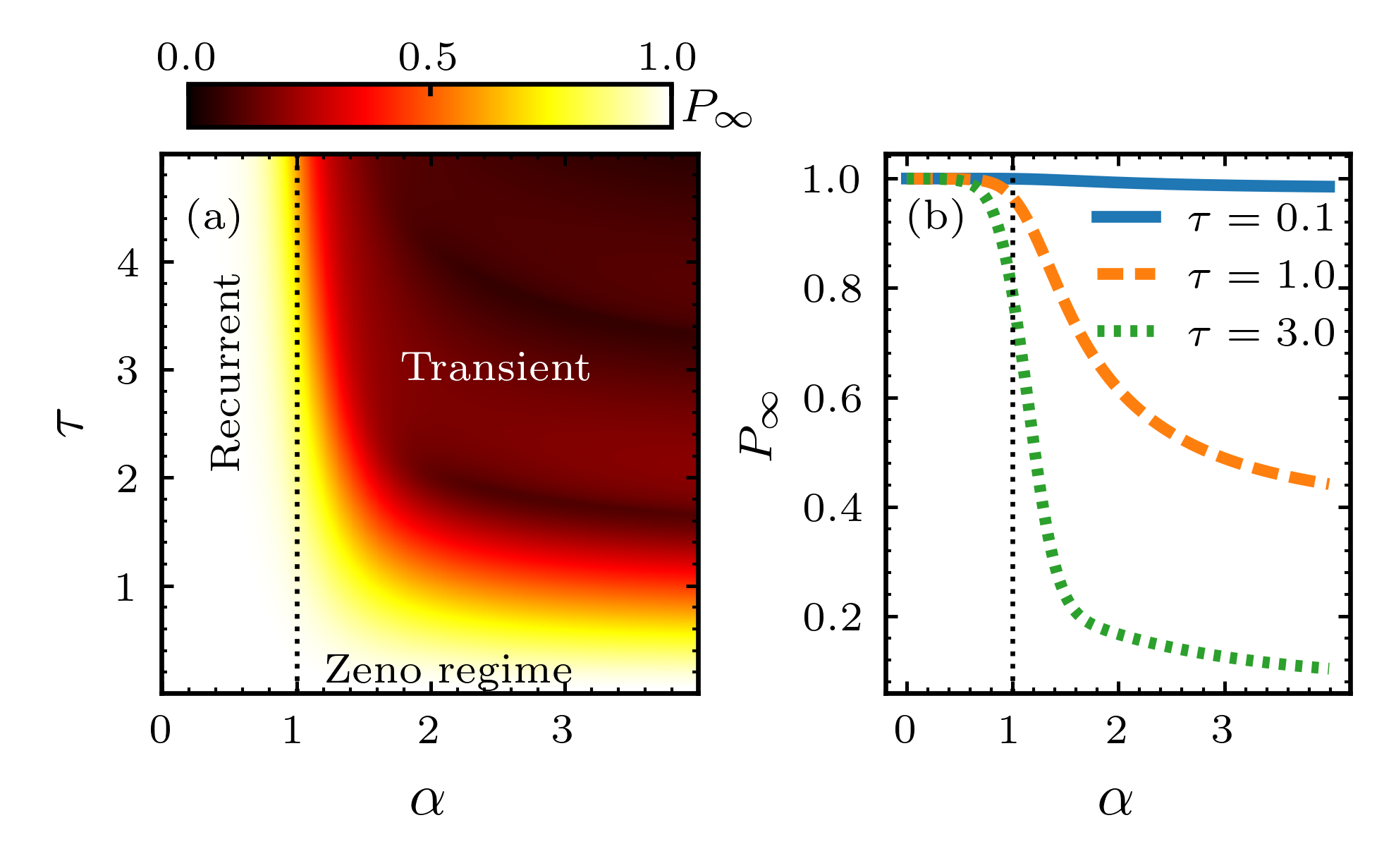}
\caption{\textit{Recurrent to transient transition.} (a) Density plot of the total detection probability $P_\infty = \sum_{n\geq 1}F_n$ as a function of hopping exponent $\alpha$ and stroboscopic time $\tau$ (in units of $1/\gamma$). The black dotted vertical line denotes the phase boundary at $\alpha=1$ between the recurrent and the transient phase; (b) $P_\infty$ as a function of $\alpha$ for $N = 10^4$ and various values of the probing time $\tau$.} 
    \label{fig:2}
\end{figure}

\begin{figure*}[!htpb]
\centering
\includegraphics[width=\textwidth]{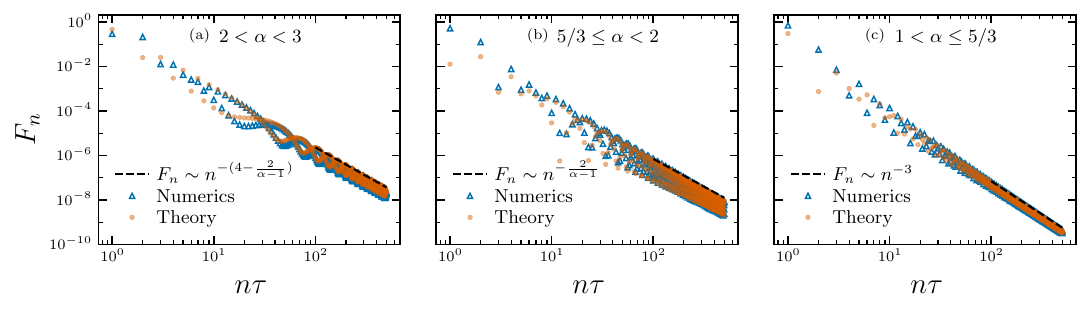}
\caption{\textit{Long-time algebraic decay in the transient phase}. Log-log plot of the first-detection probability $F_n$ as a function of number of measurement $n$. 
The panels  correspond to different regimes of hopping exponent (a) $2 < \alpha < 3$ with $\alpha = 2.4$, (b) $5/3\leq \alpha < 2$ with $\alpha = 1.8$ and (c) $1 < \alpha \leq 5/3$ with $\alpha = 1.5$. Blue triangles represent the result from exact numerical evaluations based on the quantum renewal equation with $N= 10^6, \gamma = 1, \tau = 1$. Orange circles represent the corresponding theoretical predictions for the long-time asymptotics. Black dashed lines are obtained from numerical fitting according to the power-law model indicated in the corresponding panel.} 
\label{fig:fig3}
\end{figure*}
Equation \eqref{eq:Loschmidt_asymp_main_text} is a key result of this work. Together with the nonlinear relation in Eq.~\eqref{eq:phiz}, it shows that measurement back-action leads to interference between the ultraviolet and infrared contributions. This contrasts with unitary dynamics, where the infrared contribution is subleading at long times. As we will show, this leads to the dependence of the first-detection decay exponent $\beta$ on $\alpha$, see Fig.~\ref{fig:1}(c). From Eq.~\eqref{eq:Loschmidt_asymp_main_text} one can further extract the spectral dimension $d_s$ \cite{spectral_dim_1,spectral_dim_2,King2025_optimalsearch}. The latter is related to the behavior of the energy density of states $\rho(E)$ near singular energy values $E^{\star}$: $\rho(E)\sim \vert E-E^{\star}\vert^{d_s/2-1}$. By Fourier transform, this shall correspond to the asymptotic behavior $\mathcal{L}_n\sim n^{-d_s/2}$ of the Loschmidt amplitude. In our case, we find two values of the spectral dimension, one associated with the infrared singularity, $d_s(k=0)=2/(\alpha-1)$, the other with the ultraviolet, $d_s(k=\pi)=1$. We will discuss the implications of this result later on.

\emph{Recurrence-to-transience transition.---} A natural order parameter for the recurrence-to-transience transition is the total detection probability $P_\infty$, defined as
\begin{equation}
P_\infty = \sum_{n\geq 1}F_n.
\label{eq:total_det_prob_main}
\end{equation}
For a recurrent walk, $P_\infty$ equals unity. Otherwise, if $P_\infty<1$, the walk is transient. Figure~\ref{fig:2} displays $P_\infty$ as a function of $\alpha$ and $\tau$.  In Fig.~\ref{fig:2}(a), one observes a transition from recurrent to transient at the critical tunnelling exponent $\alpha=1$. The small deviation from perfect recurrence $P_{\infty}<1$ visible in Fig.~\ref{fig:2}(b) near the transition point $\alpha = 1$ is a finite-size effect; increasing $N$ progressively suppresses this transient region \cite{SM}. The transition at $\alpha=1$ originates from the behavior of the energy gap between the ground and the first excited state of Hamiltonian \eqref{eq:Hamiltonian_main}: The gap is finite for $\alpha<1$, where the walk is recurrent, it closes logarithmically with the system size at the critical value $\alpha = 1$, and vanishes for $\alpha > 1$. From the nonanalytic behavior of $dP_{\infty}/d\alpha$ as $N\to\infty$ we extract that the transition is continuous. 
Figure \ref{fig:2} also shows that, for $\alpha>1$, the nature of the walk can change from transient to recurrent as the stroboscopic time $\tau\to 0$. This is a quantum Zeno effect, where the measurement effectively freezes the walker \cite{zeno1977one,zeno1977two}. The transition as a function of $\tau$ has the features of a crossover. We now discuss in detail the features of the individual regimes.  

\emph{Recurrent phase.--} The behavior $P_\infty=1$ implies that the walker is always detected at the initial site with unit probability. In the model, it originates from the behavior of \eqref{eq:phiz} and \eqref{eq:strong_LR_loschmidt} for $\alpha<1$, giving that $\tilde{\phi}(z) = z$, and hence $F_n = \delta_{n,1}$.
This localization has been also observed in \cite{roy_causality_2025} and emerges from the discrete nature of the energy spectrum, so that the initial state is effectively protected from spreading in the thermodynamic limit. The phenomenon exhibits the characteristic features of cooperative shielding observed in the unitary dynamics of quantum walks with all-to-all hopping \cite{Santos2016,Celardo2016}.

\emph{Transient phase.--} For $\alpha>1$, the total detection probability $P_{\infty}$ decreases monotonically with $\alpha$: the initial state is no longer protected by the shielding mechanism, and the unitary dynamics allows the walker to spread. The speed at which the walker spreads is bound by the Lieb-Robinson velocity, which decreases with $\alpha$ till reaching a saturation value at $\alpha>3$ \cite{roy_causality_2025}. 
Behind these general features, there is a subtle interplay of measurement and coherent quantum dynamics that gives rise to different scaling of the FDRP $F_n$ at long times ($n\gg 1$). For $\alpha>3$,  the infrared $k=0$ and ultraviolet $k=\pi$ modes have identical spectral dimension $d_s=1$. The model is thus effectively equivalent to a one-dimensional chain with local hopping and we recover the result $F_n \sim n^{-3}$ of Refs.~\cite{Dhar2015QuantumModel,dhar2015b,Barkai2016,Friedman2017QuantumProblem,Barkai2018}. The same scaling is found for $1<\alpha<3$ at resonant values of the probing time $\tau_{\rm{res}} = 2\pi m/(E_0 - E_\pi)$ with $m \in \mathbb{N}$. In this case $z_0$ and $z_{\pi}$ singularities in Eq.~\eqref{eq:lz_polylog_1_alpha_3_main} match and the asymptotics is governed by the ultraviolet mode $k = \pi$, giving $F_n \sim n^{-3}$ as in short-range quantum walks~\cite{SM}. Away from these resonances, the behavior of the FDRP becomes non-trivial due to the interference of infrared $z_0$ and ultraviolet $z_{\pi}$ singularities. This is captured by the exponent $\beta(\alpha)$ characterizing the asymptotic decay, $F_n\sim n^{-\beta(\alpha)}$, as summarized in Table \ref{tab:scaling_exponent}. Figure \ref{fig:1}(c) shows that the exponent $\beta$ is not a monotonous function of $\alpha$.

\begin{table}[!htpb]
    \caption{Scaling  of the algebraic tails of FDRP $F_n$ as a function of the tunneling exponent $\alpha$ for non-resonant $\tau$.} 
    \centering
    \setlength{\tabcolsep}{4pt}
    \begin{tabular}{c|c|c}
    \hline\hline
     Regime & $\alpha$ &$F_n$\\
      \hline
      (A)& $\alpha>3$&  $n^{-3}$ \\
      (B)& $2 < \alpha <3$&  $n^{-(4 - 2f(\alpha))}$\\
      (C)& $5/3 \leq \alpha < 2$&  $n^{-2 f(\alpha)}$\\
      (D)& $1< \alpha \leq 5/3$&  $n^{-3}$\\
      \hline
    \end{tabular}
    \label{tab:scaling_exponent}
\end{table}

\emph{Critical algebraic decay.---} Figure~\ref{fig:fig3} shows the behavior of the FDRP $F_n$ as a function of $n$ for  three values of $\alpha$ representative of the characteristic regimes (B), (C), and (D) of table \ref{tab:scaling_exponent}. The figure reports exact numerical evaluation of the renewal equation \eqref{eq:renewal} with the asymptotic result $F_n \sim n^{-\beta}$, obtained analytically via integration in the complex plane starting from Eqs.~\eqref{eq:inv_z_phi_main} and \eqref{eq:lz_polylog_1_alpha_3_main}, see SM~\cite{SM, ComplexPK}. 

In regime (B), Fig.~\ref{fig:fig3}(a), the first-detection probability decays as $F_n\sim n^{-\left(4- 2/(\alpha -1)\right)}$. In fact, the infrared singularity at $k=0$ dominates the generating function \eqref{eq:phiz}, and the spectral dimension is $d_s=2/(\alpha-1)$: the long-range walk on a chain is equivalent to a short-range walk in a hypercube of $d_s$ dimensions, see also~\cite{Thiel_spectral}. Note that the decay exponent $\beta(\alpha)$ is double the one of the corresponding classical Lévy flight \cite{levy_th_2}. For regimes (C) and (D), the quantum model is profoundly different from its classical Lévy-flight counterpart. In the latter, non-local jumps dominate for $1<\alpha<2$, so that the particle always overshoots the target site and the corresponding first-hitting probability vanishes \cite{levy_th_3,levy_th_5}. Quantum dynamics, instead, stabilizes a finite FDRP.

The non-monotonic behavior of $\beta$ with $\alpha$ is characterized by singular points, visible in Fig.~\ref{fig:1}(c). At $\alpha=2$ the decay exponent is minimal: the polylogarithm singularity is logarithmic, yielding $F_n\sim 1/n^2(\log n)^4$. The exponent $\alpha=3$ separates a region where $\beta$ increases (B) to the saturation value, (A). Here, the nonanalytic infrared contribution acquires a logarithmic correction and the FDRP accordingly decays as $F_n \sim \log n / n^3$~\cite{SM, Corless1996}. The critical value $\alpha=5/3$ separating regimes (C) and (D) is of different nature: at this point, the contributions of the infrared mode $F_n \sim n^{-2/(\alpha-1)}$ and of the ultraviolet one $F_n \sim n^{-3}$ are equal. At $\alpha=5/3$ the algebraic decay thus transitions from that of a long-range theory, in Fig.~\ref{fig:fig3}(b), to the one of an effectively local model with dimension  $d_s(k=\pi)=1$. This transition is caused by the measurement back-action, which determines the nonlinear relation in Eqs.~\eqref{eq:renewal} and \eqref{eq:phiz} between the first-detection return amplitude and the measurement-free Loschmidt amplitude.

\emph{Conclusion.---} Long-range hopping enhances the probability of first return of quantum walks compared to both the short-range quantum case, and the classical long-range Lévy-flight. It is responsible for the transition from recurrence to transience, which in one dimension occurs for $\alpha=1$.
The role of the measurement becomes critical in the transient regime, for $\alpha>1$: the FDRP decays algebraically with an exponent featuring multiple nonanalytic points. Such singular points are not only rooted in the infrared singularities of the energy spectrum, but they also reflect the competition between long and short-wavelength modes due to projective measurements. While measurement-induced phase transitions have so far been primarily characterized through entanglement or purification dynamics in monitored many-body systems \cite{Skinner2019,Li2019}, our results indicate that a conceptually analogous phenomenon of a measurement-driven reorganization of the effective low-energy description can already emerge in the single-particle sector. Here, the interplay between coherent long-range spreading and stroboscopic projective measurement generates nonanalytic signatures in the first-detection statistics that have no counterpart in the unitary dynamics alone.
The different dynamical regimes observed by us could be unveiled in experimental platforms which permit to tune the long-range exponent $\alpha$, such as trapped ions \cite{trapped_ions_1,trapped_ions_2}, neutral atoms in photonic modes \cite{cavity_1,cavity_2} and Rydberg atoms \cite{rydb1,rydb2,rydb3}. 

\emph{Data availability statement.---}The data that support the findings of this article are publicly available on Zenodo at \cite{data}.

\emph{Acknowledgments.---} The authors acknowledge fruitful discussions with Emma King and Francesco Mattiotti. This work was funded by the German Ministry of Education and Research (BMBF, Project ``NiQ: Noise in Quantum Algorithms"), by the Deutsche Forschungsgemeinschaft (DFG, German Research Foundation) – Project-ID 429529648 – TRR 306 QuCoLiMa (``Quantum Cooperativity of Light and Matter''), and by the QuantERA II Programme (project ``QNet: Quantum transport, metastability, and neuromorphic applications in Quantum Networks"), which has received funding from the EU's Horizon 2020 research and innovation programme under Grant Agreement No.\ 101017733, as well as from the Deutsche Forschungsgemeinschaft DFG (Project ID 532771420). SG acknowledges Alexander von Humboldt Foundation for support under its Humboldt Research Fellowship for Experienced Researchers. SG also acknowledges generous allocation of computational resources of the Department of Theoretical
Physics, TIFR, assistance of Kapil Ghadiali and Ajay Salve, and the financial support
of the Department of Atomic Energy, Government of India under Project Identification
No. RTI 4002. G.P. acknowledges funding from the Swiss National Science Foundation (SNSF) through Grant No.~10005336.

\makeatletter
\let\originaladdcontentsline\addcontentsline
\renewcommand{\addcontentsline}[3]{}
\bibliography{refs}
\let\addcontentsline\originaladdcontentsline
\makeatother

\setcounter{equation}{0}
\setcounter{figure}{0}
\setcounter{table}{0}
\renewcommand{\theequation}{S\arabic{equation}}
\renewcommand{\thefigure}{S\arabic{figure}}

\makeatletter
\renewcommand{\theequation}{S\arabic{figure}}
\renewcommand{\thefigure}{S\arabic{figure}}

\onecolumngrid
\newpage

\setcounter{page}{1}

\setcounter{secnumdepth}{3}
\pagestyle{plain}
\begin{center}
{\Large SUPPLEMENTAL MATERIAL}
\end{center}
\begin{center}
\vspace{0.8cm}
{\Large Phase transitions in first-detection statistics of monitored long-range quantum walks}
\end{center}
\begin{center}
Sayan Roy$^1$, Shamik Gupta$^2$, Giovanna Morigi$^{1,3}$, and Gabriele Perfetto$^4$
\end{center}
\begin{center}
$^1${\em Theoretische Physik, Universit\"{a}t des Saarlandes, D-66123 Saarbr\"{u}cken, Germany}\\
$^2${\em Department of Theoretical Physics, Tata Institute of Fundamental Research, Homi Bhabha Road, Mumbai 400005, India.}\\
$^3${\em Center for Quantum Technologies (QuTe), Saarland University, Campus, 66123 Saarbr\"ucken, Germany}\\
$^4${\em Institut für Theoretische Physik, ETH Zürich, Wolfgang-Pauli-Str. 27, 8093 Zürich, Switzerland}
\end{center}
\setcounter{equation}{0}
\setcounter{figure}{0}
\setcounter{table}{0}
\setcounter{page}{1}
\makeatletter
\renewcommand{\theequation}{S\arabic{equation}}
\renewcommand{\thefigure}{S\arabic{figure}}
\makeatletter
\renewcommand{\theequation}{S\arabic{equation}}
\renewcommand{\thefigure}{S\arabic{figure}}
\renewcommand{\bibnumfmt}[1]{[S#1]}
\renewcommand{\citenumfont}[1]{S#1}
\onecolumngrid
\setcounter{secnumdepth}{3}

\tableofcontents

\section{Eigenvalues and eigenvectors of the one-dimensional long-range Hamiltonian}
\label{app:I}
We consider a single quantum walker hopping between the sites on a one-dimensional lattice of $N$ sites. The unitary evolution of the walker is dictated by the Hamiltonian $\hat{H}_\alpha$ given by
\begin{align}
    \hat{H}_\alpha = - \frac{\gamma}{\mathcal{N}_\alpha}\sum_{i < j} \frac{1}{d_{ij}^\alpha} \left(\ket{i}\bra{j} + \mathrm{H.c.}\right)\,,
    \label{eq:Hamiltonian}
\end{align}
where $|i\rangle$ denotes the state of the walker on site $i$ ($i=0,\ldots,N-1$), $d_{ij}$ the dimensionless distance between sites $i$ and $j$, and the constant $\gamma >0$ with the dimension of energy characterizing the tunneling probability between a pair of sites $i$ and $j$. The exponent $\alpha$ characterizes the decay of the tunneling probability with distance. The limit $\alpha\to\infty$ corresponds to nearest-neighbor hopping, while finite $\alpha$ values allow for long-range tunneling. We assume periodic boundary conditions, and hence, $d_{ij}\equiv \mathrm{min}(|i-j|,N-|i-j|)$ is the minimum distance between sites $i$ and $j$ on the lattice with lattice spacing set to unity. We also work in units such that $\hbar=1$.

To obtain a well-defined thermodynamic limit for long-range interactions, we employ the Kac's prescription, whereby the factor $\mathcal{N}_\alpha = \sum_{d = 1}^{\lfloor N/2 \rfloor} d^{- \alpha}$ has to be introduced~\cite{Kastner2011DivergingModels, kastner2025longrangesystemsnonextensivityrescaling}. For large $N$, one has the scaling
\begin{align}
  \mathcal{N}_\alpha  \approx  \begin{cases}
			\frac{2^{\alpha - 1}}{1- \alpha} N^{1- \alpha} & \text{if $\alpha  < 1$}\,,\\
            \log N & \text{if $\alpha = 1$}\,, \\
             \zeta(\alpha) & \text{if $\alpha > 1$}\,,
		 \end{cases}
         \label{eq:kac_scaling}
\end{align}
where $\zeta(\alpha)$ is the Riemann Zeta function.


\subsection{Nearest-neighbor case ($\alpha \to \infty$)}
\label{app:IA}
We first recall the  nearest-neighbor limit, in which the Hamiltonian reads
\begin{align}
    \hat{H}_{\rm NN} = - \gamma\sum_{i} \left(\ket{i}\bra{i+1} +\rm{H.c.} \right),
    \label{eq:Hamiltonian_NN}
\end{align}
where we have used $\mathcal{N}_\infty= 1$. Using translational invariance under periodic boundary conditions, the eigenvectors are the discrete  Fourier modes

\begin{align}
      \ket{k_l}= \frac{1}{\sqrt{N}}\sum_{m=0}^{N-1} e^{ik_l m}\ket{m} = \frac{1}{\sqrt{N}}\sum_{m=0}^{N-1} c_m\ket{m}, \quad c_m=e^{i k_l m},
    \label{eq:fourier_modes}
\end{align}
with $k_l=\frac{2\pi l}{N},\quad l=0,1,\ldots,N-1.$ Using the eigenvalue equation $\hat{H}_{\rm{NN}}\ket{k_l} = E_{k_l} \ket{k_l}$, one finds the eigenvalues 
\begin{align}
    E_{k_l} = - 2 \gamma \cos(k_l)\,.
\end{align}
 For even $N$, except for the ground state ($k_l = 0$) and the highest excited state ($k_l = \pi$), all the other states are two-fold degenerate, which leads to $(N + 2)/2$ distinct eigenvalues. For odd $N$, except for the ground state, all the other states are two-fold degenerate, which corresponds to $(N + 1)/2$ distinct eigenvalues.


\subsection{Long-range case}
\label{app:IB}
The long-range Hamiltonian is translationally invariant and therefore has a circulant matrix representation in the site basis. Consequently, it is diagonalized by the same discrete Fourier modes as the nearest-neighbor Hamiltonian. For an odd number $N$ of lattice sites, the Hamiltonian takes the form
\begin{align}
     \hat H_{\alpha} = - \frac{\gamma}{\mathcal{N}_\alpha} \sum_{m = 0}^{N-1} \sum_{d = 1}^{\frac{N - 1}{2}} \frac{1}{d^\alpha} \left(\ket{m}\bra{m + d} + \rm{H.c.} \right)\,,
\end{align}
For an even number $N$ of lattice sites, we have to be careful about the double counting of the middle sites, and one has
\begin{align}
     \hat{H}_{\alpha} = - \frac{\gamma}{\mathcal{N}_\alpha} \sum_{m = 0}^{N-1} \sum_{d = 1}^{\frac{N}{2} - 1} \frac{1}{d^\alpha} \left(\ket{m}\bra{m + d} + \rm{H.c.} \right) - \hat H_{\rm even}\,,
\end{align}
with 
\begin{align}
  \hat H_{\rm even} = \frac{\gamma}{2 \mathcal{N}_\alpha}  \sum_{m = 0}^{N-1} \left(\frac{N}{2}\right)^{-\alpha} \left(\ket{m}\bra{m + N/2} + \rm{H.c.} \right)\,.
    \label{eq:H_even}
\end{align}
Taking the inner product of the eigenvalue equation $\hat{H}_{\alpha}\ket{k_l} = E_{k_l}(\alpha) \ket{k_l}$  with $\bra{m^\prime}$ on both sides for any general $\alpha$, one finds 
\begin{align}
    E_{k_l}(\alpha) c_{m^\prime} =   - \frac{\gamma}{\mathcal{N}_\alpha}  \sum_{d = 1}^{\frac{N - 1}{2}} \frac{1}{d^\alpha} \left(c_{m^\prime+d} + c_{m^\prime-d} \right)
   \implies E_{k_l}(\alpha)  =  -  \frac{2\gamma}{\mathcal{N}_\alpha}  \sum_{d = 1}^{\frac{N - 1}{2}} \frac{1}{d^\alpha} \cos(k_l d)\,.   
   \label{eq:E_alpha_odd}
\end{align}
For the even $N$ case, we have to consider the contribution from the extra term $\hat H_{\rm even}$ as stated in Eq.~\eqref{eq:H_even}. Following the same procedure as used for Eq.~\eqref{eq:E_alpha_odd}, we get
\begin{align}
   E_{k_l}^{\rm even}(\alpha) = \frac{\gamma}{\mathcal{N}_\alpha}  \left(\frac{N}{2}\right)^{-\alpha} (-1)^l\,.
\end{align}
Thus, the eigenvalues for a periodic chain with long-range hopping are
\begin{align}
E_{k_l}(\alpha) =
\begin{cases}
\displaystyle
- \frac{2\gamma}{\mathcal{N}_\alpha}
\sum_{d = 1}^{\frac{N - 1}{2}}\frac{1}{d^\alpha}\cos(k_l d),& \text{$N$ odd}\,, \\[2ex]
\displaystyle
- \frac{2\gamma}{\mathcal{N}_\alpha}\sum_{d = 1}^{\frac{N}{2} - 1}\frac{1}{d^\alpha}\cos(k_ld) -  E_{k_l}^{\rm even}(\alpha) ,& \text{$N$ even}\,.
\end{cases}
\label{eq:energy_spectrum}
\end{align}

\subsection{Spectrum in the thermodynamic limit}
\label{app:IC}
In this section, we discuss the main spectral features in the thermodynamic limit already obtained in Ref.~\cite{Defenu2021MetastabilitySystems, defenu_out--equilibrium_2024}. The additional correction term $E_{k_l}^{\rm even}(\alpha)$ for even $N$ scales as
\begin{align}
\vert E_{k_l}^{\rm even}(\alpha) \vert
\sim
\begin{cases}
N^{-1}, & \alpha < 1\,, \\[0.5ex]
(N\log N)^{-1}, & \alpha = 1\,, \\[0.5ex]
N^{-\alpha}, & \alpha > 1\,,
\end{cases}
\end{align}
which vanishes in the thermodynamic limit, and the spectra for odd and even $N$ become equivalent. The single-particle spectrum can then be written as
\begin{align}
E_{k_l}(\alpha)= -\frac{2\gamma}{\mathcal{N}_\alpha}\sum_{d\geq1}\frac{\cos(k_ld)}{d^\alpha}\,,
\label{eq:energy_spectrum_TL}
\end{align}
 For $\alpha>1$, in the weak long-range regime, the Kac's normalization $\mathcal{N}_\alpha$ remains finite in the thermodynamic limit $N\to \infty$ and one can take the thermodynamic limit in a standard way and proceed as one does for the nearest-neighbor case. The allowed discrete momentum values $k_l = 2\pi l/N$ become dense in the Brillouin zone and can be replaced with the continuous variable $k \in [0,2\pi)$. Accordingly, the spectrum becomes also continuous.

In contrast, for $0 < \alpha < 1$, in the strong long-range regime, one needs to take the thermodynamic limit a bit more carefully since the Kac's normalization $\mathcal{N}_\alpha$ diverges for large $N$ and grows like $N^{1-\alpha}$ according to Eq.~\eqref{eq:kac_scaling}. In this regime, the values of $E_{k}(\alpha)$ progressively squeeze towards an inverted delta function with minimum at $k =0$ as $N \to \infty$; upon zooming near $k = 0$, one finds a sequence of discrete finite values. To see this more concretely, we first consider a discrete set of momenta $k_l = 2 \pi l/N$ and then perform a change of variable $x = d/N$ in Eq.~\eqref{eq:energy_spectrum_TL} and then recast the summation in $d$ to integration in $x$ which results in~\cite{Defenu2021MetastabilitySystems, defenu_out--equilibrium_2024}
\begin{align}\label{eq:E_alpha_less_1}
    E_l(\alpha) = - \frac{2 \gamma(1 - \alpha)}{2^{\alpha - 1}} \displaystyle\int_0^{1/2} \mathrm{d}x \, x^{-\alpha} \cos(2 \pi l x)\,.
\end{align}
Note that the energy eigenvalues depend only on the discrete integer index $l \in \mathbb{Z}$ rather than on the continuous momentum $k$. For mode index $l=0$, it gives $E_0 = -2\gamma$ independent of $\alpha$. For $l > 0$, changing variable to $u = 2 \pi l x = k_lN x$, gives the energy as $E_{k_l}(\alpha) = -2\gamma(1-\alpha)2^{1-\alpha} I(k_lN)$, with 
\begin{align}\label{eq:integral_spectrum}
    I(k_lN) = (k_lN)^{\alpha-1}\int_0^{k_lN/2}du\,u^{-\alpha}\cos u\,.
\end{align}
 If $l=\mathcal{O}(1)$ implying $k_l N = 2 \pi l = \mathcal{O}(1)$, the integral~\eqref{eq:integral_spectrum} becomes $N$ independent and finite. Thus $E_{k_l}$ has a finite thermodynamic limit for $l = \mathcal{O}(1)$ and these finite-energy levels form a discrete sequence labeled by $l$, rather than a continuous dispersion
in $k$ as shown in Eq.~\eqref{eq:E_alpha_less_1}. If instead one takes the limit of $l \to \infty$ in the thermodynamic limit spectrum, one has $k_l N = 2 \pi l \to\infty$, and hence
\begin{align}
I(k_l N) \sim (k_lN)^{\alpha-1}
\int_0^\infty du\,u^{-\alpha}\cos u = \Gamma(1-\alpha)\sin\left(\frac{\pi\alpha}{2}\right) (k_lN)^{\alpha-1}\,.
\end{align}
Since $\alpha-1<0$, this vanishes as $(k_l N)^{\alpha-1}$.  This implies in the thermodynamic limit, for long-wavelength $k = 0$ mode there is a discrete set of finite energy eigenvalues squeezed on the vertical axis. For finite momentum $k \neq 0$ modes, the energy, instead, asymptotically vanishes as $N\to \infty$.  

\begin{figure*}[!htpb]
\centering
\includegraphics[width=\textwidth]{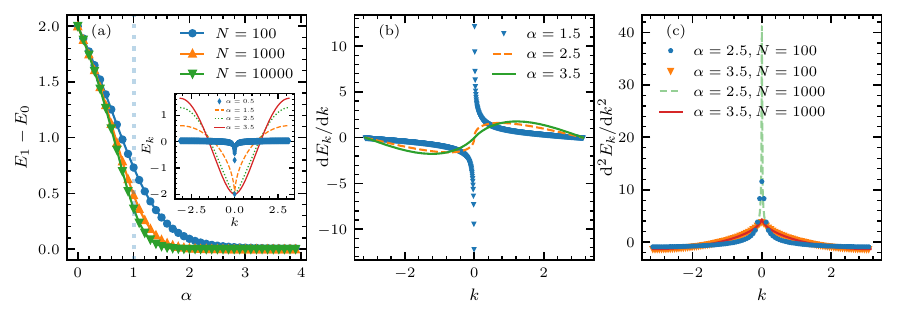}
\caption{\textit{Properties of the energy spectrum for various $\alpha$ values.} (a) Spectral gap between the first excited state $E_1$ and the ground state $E_0$ as a function of the hopping exponent $\alpha$ for three different system sizes $N$. The vertical dotted line represents $\alpha= 1$  which demarcates the transition from finite gap to gapless spectrum in the thermodynamic limit. (Inset). Dispersion relation $E_k$ as a function of $k$ for different $\alpha$. Other parameters are $N = 1000,\, \gamma = 1$. (b) First derivative of the energy spectrum $ \mathrm{d}E_k/\mathrm{d}k$ as a function of $k$ for three representative $\alpha$ values. It diverges for $1< \alpha< 2$ at $k = 0$ which is represented by $\alpha = 1.5$ in blue triangles. Otherwise the derivative is finite for $\alpha >2$. Other parameters are $N = 1000,\, \gamma = 1$. (c) Second derivative of the energy spectrum $ \mathrm{d}^2E_k/\mathrm{d}k^2$ as a function of $k$ for two different $N$ values. This quantity diverges at $k = 0$ for increasing $N$ for $2 <\alpha < 3$, represented by $\alpha = 2.5$, whereas it remains finite for $\alpha >3$, represented by $\alpha = 3.5$.} 
\label{fig:figS1}
\end{figure*}

In Fig.~\ref{fig:figS1}(a), we plot numerically Eq.~\eqref{eq:energy_spectrum} to show how the energy spectrum behavior changes for various values of the tunneling exponent $\alpha$ for increasing $N$. In Fig.~\ref{fig:figS1}(a), we, namely plot the energy gap $E_1-E_0$ between the bottom of the energy band $k_{l=0}$ and the first excited state with $k_{l=1}$. One can see that in the thermodynamic limit $N\to \infty$, the gap closes for $\alpha>1$, while it remains finite for $\alpha<1$. In Figs.~\ref{fig:figS1}(b) and (c), we plot the first derivative of the energy spectrum (group velocity) and the second one. One can see that the group velocity becomes singular at $k=0$ for $1 < \alpha < 2$, while the second derivative becomes singular at $k=0$ for $2 < \alpha < 3$. The integer values of $\alpha = 0, 1,2$ and $3$ are special cases which are discussed separately in Subsecs.~\ref{app:IIIA}, \ref{app:IIIC}, \ref{app:IIIE} and  \ref{app:IIIG}, respectively.

\section{Renewal Formalism and connection to Loschmidt Amplitude}
\label{app:II}
In the stroboscopic measurement protocol discussed in the main text, the quantity of interest is the first-detection return amplitude $\phi_n$. Introducing the time-evolution operator $\hat U(\tau) = e^{-i \hat H \tau}$, $\phi_n$ is known to satisfy the quantum renewal equation~\cite{Dhar2015QuantumModel,Friedman2017QuantumProblem, Walter2025ThermodynamicSystems,roy_causality_2025}:
\begin{align}
    \phi_n
    = \bra{0}\hat{U}(n\tau)\ket{0}
    -\sum_{j=1}^{n-1}\phi_j\,\bra{0}\hat{U}\!\bigl((n-j)\tau\bigr)\ket{0}\,.
    \label{eq:renewal_SM}
\end{align}
The corresponding first-detection return probability is $F_n = |\phi_n|^2$. In terms of $z$-transforms, $ \tilde{\phi}(z)=\sum_{n\geq1} z^{n}\phi_n$ of $\phi_n$ and $\tilde{U}(z)=\sum_{n\geq 1} z^{n}\hat{U}(n\tau)$ of the unitary operator $\hat U(\tau)$, Eq.~\eqref{eq:renewal_SM} becomes
\begin{align}
    \tilde{\phi}(z)
    =\frac{\bra{0}\tilde{U}(z)\ket{0}}{1+\bra{0}\tilde{U}(z)\ket{0}}.
    \label{eq:phi_tilde_solution}
\end{align}
$\tilde{\phi}(z)$ is obtained by solving the above equation which yields the $\phi_n$'s by performing the inverse $z$-transform,
\begin{align}
    \phi_n=\frac{1}{2\pi i}\oint_C dz\,\frac{\tilde{\phi}(z)}{z^{n+1}},
    \label{eq:inv_z_phi}
\end{align}
where $C$ is a contour encircling the origin within the domain of analyticity of $\tilde{\phi}(z)$. We write Eq.~\eqref{eq:phi_tilde_solution} in terms of the $z$-transform of the Loschmidt amplitude
\begin{align}
\mathcal{L}_n \equiv \bra{0} \hat{U}(n\tau)\ket{0} = \frac{1}{N} \sum_{l = 0}^{N -1}\, \exp(-i E_{k_l}(\alpha) n\tau). \label{eq:L}
\end{align}
The squared modulus of this quantity gives the probability that the walker returns to its initial site $\ket{0}$ after a time $n\tau$ according to the measurement-free unitary dynamics. One then has
 \begin{align}
    \tilde{\phi}(z)=\frac{\tilde{\mathcal{L}}(z)}{1+\tilde{\mathcal{L}}(z)}\,.
    \label{eq:phi_L_relation}
\end{align}
Equation~\eqref{eq:phi_L_relation} shows that the long-time behavior of $\phi_n$ (and hence of $F_n$) is governed by the singularity structure of $\tilde{\mathcal{L}}(z)$ in the complex-$z$ plane. In the following subsections, we determine $\tilde{\mathcal{L}}(z)$ in the thermodynamic limit and identify its leading nonanalyticities as a function of the hopping exponent $\alpha$, from which the asymptotic first-detection behavior follows via Eq.~\eqref{eq:inv_z_phi}. 


\section{Detailed Calculation of Loschmidt Amplitude $\mathcal{L}_n$ for different $\alpha$}
\label{app:III}


\subsection{Case for $\alpha = 0$}
\label{app:IIIA}
This is the fully-connected case, where the quantum particle can hop between any two sites with equal probability. Owing to the permutation symmetry, the energy spectrum \eqref{eq:energy_spectrum} reduces to two energy levels: a ground state with eigenenergy $-2\gamma(N -1)/N$ and an excited state with eigenenergy $2\gamma/N$ which is $(N-1)$-fold degenerate. Equation~\eqref{eq:L} then gives
\begin{align}
    \mathcal{L}_n = \frac{N-1}{N} e^{-i \frac{2\gamma n\tau}{N}} + \frac{1}{N} e^{i \frac{2\gamma(N -1) n\tau}{N}} =  e^{-i \frac{2\gamma n\tau}{N}} \left[ 1 - \mathcal{O}\left(\frac{1}{N}\right)\right],
\end{align}
which in the thermodynamic limit reduces to $\mathcal{L}_n = e^{-i \frac{2\gamma n\tau}{N}} \to 1$.


\subsection{Case for $0<\alpha <1$}
\label{app:IIIB}

This is the strong long-range case. The spectrum remains discrete even in the thermodynamic limit, as discussed in Sec.~\ref{app:IC}. Using the energy spectrum expression in Eq.~\eqref{eq:energy_spectrum}, we can find for a given $N$ a suitable value $M \sim \mathcal{O}(1)$ such that $E_{k_l}(\alpha)$ is arbitrarily small for $l > M$. Consequently, we have
\begin{align}\label{eq:ln_expand_alpha<1}
    \mathcal{L}_n =  \frac{1}{N} \sum_{l = 0}^{M}\, \exp(-i E_{k_l}(\alpha) n\tau) + \frac{1}{N} \sum_{l = M+1}^{N-1}\, \exp(-i E_{k_l}(\alpha) n\tau)\,.
\end{align}
In the thermodynamic limit, the first term that accounts for long-wavelength modes goes to $0$ since one can bound $\sum_{l = 0}^{M}\, |\exp(-i E_{k_l}(\alpha) n\tau)| \leq M$, and the ratio $M/N \to 0$. The non-vanishing contribution comes only from the second part describing short wavelengths, which have energies $E_{k_l} = 0$ in the thermodynamic limit. It yields $\frac{1}{N} \sum_{l = M+1}^{N-1}\, \exp(-i E_{k_l}(\alpha) n\tau) = 1 - \mathcal{O}(1/N)$, which consequently leads to $\mathcal{L}_n \to 1$. This result, together with the analysis of Subsec.~\ref{app:IIIA}, justifies the result in Eq.~(7) of the main text.


\subsection{Case for $\alpha = 1$}
\label{app:IIIC}
This is the transition point from the discrete nature of the spectrum $\alpha < 1$ to the continuous nature of the spectrum for $\alpha > 1$. At $\alpha = 1$, the Kac's normalization also diverges logarithmically. For modes whose index scales as $l=\mathcal{O}(N^{a})$, with $0<a<1$, one can simplify the energy spectrum as 
\begin{align}
    E_l(\alpha=1) &=- \frac{2 \gamma}{\log N} \sum_{d\geq 1} \frac{\cos (2\pi l d/N)}{d} = -  \frac{2 \gamma}{\log N}  \sum_{d\geq 1} \frac{e^{i2\pi l d/N} + e^{-i2\pi l d/N}}{2d} \nonumber \\ &= - \frac{2 \gamma}{\log N} \left[-\frac{1}{2} \log(1 - e^{i2 \pi l/N}) - \frac{1}{2} \log(1 - e^{-i2 \pi l/N})  \right] = - \frac{2 \gamma}{\log N} \left[ - \log(2 \sin\, \frac{\pi l}{N})\right].
\end{align}
Taking the limit $N \to \infty$, one has $l/N \to 0$, so $\sin\, \frac{\pi l}{N} \approx \pi l/N$. The constant factors are irrelevant after division by $\log N$. Therefore, 
\begin{align}
    E_l(1) \approx - 2 \gamma \frac{\log (N/l)}{\log N} = - 2\gamma (1 - a).
\end{align}
Modes with $l = \mathcal{O}(1)$ have energy equal to $- 2 \gamma$. Modes with $l =\mathcal{O}(N^a)$, with $0<a<1$, corresponding to momenta $k_l = \frac{2\pi l}{N} \to 0$ in the thermodynamic limit have finite energy $-2\gamma(1- a)$. In contrast, for bulk of the modes with $l =  \mathcal{O}(N)$, corresponding to finite momenta $k \neq 0$ have zero energy. Therefore, in the thermodynamic limit, a continuum of finite energies accumulates at $k = 0$. This is similar to the case $\alpha <1$, where finite-energy modes also collapsed onto $k = 0$. The important difference is that for $\alpha <1$ the energy spectrum remains discrete as $N\to \infty$, whereas at $\alpha = 1$, it becomes continuous. This shows explicitly how the marginal case $\alpha = 1$ interpolates between the two regimes. Consequently, in the thermodynamic limit, the Loschmidt amplitude is dominated by these extensive zero energy modes and one obtains $\mathcal L_n \to 1$. This result, together with the analysis of Subsecs.~\ref{app:IIIA} and \ref{app:IIIB}, justifies the result in Eq.~(7) of the main text.


\subsection{Case for $ 1< \alpha < 2$}
\label{app:IIID}
Since the spectrum $E_k(\alpha)$ is continuous in the thermodynamic limit, one can express the Loschmidt amplitude in Eq.~\eqref{eq:L} as an integral over the first Brillouin zone $k\in [0,2\pi)$:
\begin{align}\label{eq:ln_alphag1}
    \mathcal{L}_n =  \frac{1}{2 \pi} \int_{0}^{2 \pi} d k\,  e^{-i E_k(\alpha) n \tau}\,,
\end{align}
where we can recast the energy spectrum in terms of polylogarithm function as
\begin{align}
    E_k(\alpha) = -2 \gamma \frac{\rm{Re} \left[Li_\alpha(e^{ik})\right]}{\zeta(\alpha)}\,,
\end{align}
where $\mathrm{Li}_\alpha(z)$ denotes polylogarithm function of order $\alpha$ and argument $z$. The Eq.~\eqref{eq:ln_alphag1} coincides with Eq.~(8) of the main text. For large $n \tau$, the integrand is highly oscillating and integration of fast oscillating functions can be obtained by stationary phase approximations. The method tells us that the leading contribution of the integral comes from the neighborhood of the stationary points $k^*$ which satisfy $\frac{\mathrm{d}E_k}{\mathrm{d}k}\vert_{k=k^*}=0$.

For $1 < \alpha <2$, $k^*=0$ is not a stationary point because of the branch-point singularity of polylogarithm at $1$ and consequently the derivative diverges.  However, this is a point of nonanalyticity of the integrand and thus contributes to the asymptotics of both $\mathcal{L}_n$ and eventually plays a major role in the asymptotics of the first-detection amplitude $\phi_n$. To show that, we explicitly rewrite Eq.~\eqref{eq:ln_alphag1} as
\begin{align}\label{eq:Ln_detailed}
        \mathcal{L}_n =  \frac{1}{2 \pi} \int_{0}^{2 \pi} d k\,  e^{-i E_k(\alpha)t}\, = \frac{1}{ \pi} \int_{0}^{\pi} d k\,  e^{-i E_k(\alpha)t} &= \frac{1}{ \pi} \left[ \int_{0}^{\epsilon}\mathrm{d} k + \int_{\epsilon}^{\pi - \delta}\mathrm{d} k + \int_{\pi - \delta}^{\pi}\mathrm{d}k \right] e^{-i E_k(\alpha)t}\,\nonumber \\
        &\equiv I_1 +I_2 + I_3,
\end{align}
where we have denoted $t = n\tau$ and $\delta, \epsilon >0$ are infinitesimally small numbers. In the first line we have used the periodicity of the argument of the exponential, i.e., $E_k(\alpha) = E_{2\pi - k}(\alpha)$. In the second line, we have split the integral limits in the neighborhood of the singular point $k^* = 0$ and the stationary point $k^* = \pi$. In the neighborhood of $k^* = 0$, the local expansion follows
\begin{align}\label{eq:local_exp_0}
     E_k(\alpha) &\approx  -2\gamma  - 2 \gamma \frac{\Gamma(1-\alpha) \cos(\frac{\pi}{2}(\alpha -1))}{\zeta(\alpha)} |k|^{\alpha - 1} = E_0 + c_\alpha |k|^{p}\,,
\end{align}
where $E_0 = -2 \gamma$, $c_\alpha = -2 \gamma \frac{\Gamma(1-\alpha) \cos(\frac{\pi}{2}(\alpha -1))}{\zeta(\alpha)}$ and 
\begin{equation}
p = \alpha -1. 
\end{equation}
Putting this in $I_1$, one gets
\begin{align}
    I_1 &\approx \frac{e^{-it E_0}}{\pi} \int_{0}^{\epsilon} \mathrm{d} k\, e^{-itc_{\alpha}k^p} = \frac{e^{-it E_0}}{\pi p} (i c_\alpha t)^{-1/p} \left[ \Gamma\left(\frac{1}{p}\right) - \Gamma\left( \frac{1}{p}, i c_\alpha t \epsilon^p \right)   \right]\,,
\end{align}
where $\Gamma(z)$ is the Euler gamma function and $\Gamma(s,z)$ is an upper incomplete gamma function \cite{NIST:DLMF}. Now for large-$t$ asymptotics, if one assumes $t^{-1/p }\ll \epsilon \ll1$, which implies $t \epsilon^p \gg 1$, one can perform the large $z$ expansion ($|z| \to \infty$) of the upper incomplete gamma function as follows \cite{NIST:DLMF}
\begin{align}
    \Gamma(s,z) = z^{s -1} e^{-z}\left[1 + \frac{s - 1}{z} + \mathcal{O}(1/z^2) \right]\,,
\end{align}
which implies 
\begin{align}
    \Gamma\left( \frac{1}{p},  i c_\alpha t \epsilon^p \right) \approx ( i c_\alpha t \epsilon^p )^{\frac{1}{p}  - 1} e^{ -i c_\alpha t \epsilon^p } \left[ 1 -  \frac{p - 1}{i p c_\alpha t \epsilon^p } \right] \,.\nonumber\\
\end{align}
This results in
\begin{align}
\label{eq:I1}
    I_1 &= \frac{e^{-it E_0}}{\pi p} (i c_\alpha t)^{-1/p} \, \Gamma\left(\frac{1}{p}\right) - \frac{e^{-it (E_0 + c_\alpha \epsilon^p)}}{\pi }  \left[ \frac{1}{it c_\alpha p\epsilon^{p -1}} -  \frac{p - 1}{(it)^2 (p c_\alpha)^2  \epsilon^{2p-1} } + \ldots\right]\,. 
\end{align}

Now, let's consider the case of $I_2$, where the derivative is always finite and it corresponds to the bulk region. Integration by parts gives 
\begin{align}
    I_2 &= \frac{1}{\pi}\int_{\epsilon}^{\pi - \delta} \mathrm{d}k\, e^{-iE_k(\alpha) t} 
    = -\frac{e^{-i E_k(\alpha)t}}{it \pi E_k^{\prime}(\alpha) } \Bigg\vert_{k = \epsilon}^{k = \pi - \delta} + \frac{E_k^{\prime \prime}(\alpha)\,e^{-i E_k(\alpha)t}}{(it)^2 \pi [E_k^{\prime}(\alpha)]^3 } \Bigg\vert_{k = \epsilon}^{k= \pi - \delta} +\ldots\,.\label{eq:I2}
\end{align}
Now, for computing the derivative of the energy functional at $\theta = \epsilon$, we will use the local expansion as written before in Eq.~\eqref{eq:local_exp_0}. Then it follows
\begin{align}
    E_k^{\prime}(\alpha)\vert_{k= \epsilon} &= p c_\alpha \epsilon^{p -1}\,, \\
    E_k^{\prime \prime}(\alpha)\vert_{k = \epsilon} &= p(p-1) c_\alpha \epsilon^{p-2}\,,
\end{align}
This results in $ E_k^{\prime \prime}(\alpha)/[E_k^{\prime}(\alpha)]^3 = p(p-1) c_\alpha \epsilon^{p-2}/ p^3 c_\alpha^3 \epsilon^{3p - 3} = (p-1)/((pc_\alpha)^2 \epsilon^{2p -1})$. One can now clearly see  the contribution to the integral $I_1$ due to the endpoint $\epsilon$ exactly cancels out with the endpoint contribution to the integral $I_2$. 

For the case of $I_3$, we use the local expansion of $E_k(\alpha)$ near $k = \pi$ which reads as
\begin{align}\label{eq:expansion_pi}
    E_{k}(\alpha) &\approx -2 \gamma(2^{1- \alpha} -  1) - \gamma (1 - 2^{3- \alpha})\frac{\zeta(\alpha -2)}{\zeta(\alpha)} (k - \pi)^2 = E_\pi + d_\alpha (k - \pi)^2\,,
\end{align}
where $E_\pi =  -2 \gamma(2^{1- \alpha} -  1)$ and $d_\alpha = -\gamma (1 - 2^{3- \alpha})\frac{\zeta(\alpha -2)}{\zeta(\alpha)}$. Setting $x = \pi - k$, one finds $I_3$ as
\begin{align}
    I_3  = \frac{1}{\pi}\int_{\pi - \delta}^{\pi} \mathrm{d}k\, e^{-iE_k(\alpha) t} = \frac{e^{-itE_\pi}}{\pi}\int_{0}^{\delta} \mathrm{d}x\, e^{-it\,d_\alpha x^2} 
    &=  \frac{e^{-itE_\pi}}{\pi} \left[\int_{0}^{\infty} \mathrm{d}x -  \int_{\delta}^{\infty}  \mathrm{d}x    \right ]\, e^{-it\,d_\alpha x^2} \nonumber \\
    &= I_3^{\rm(sp)} - I_3^{\rm(corr)}.
\end{align}
The first integral is the main stationary phase which forms the standard Gaussian integral and follows
\begin{align}
    I_3^{\rm(sp)} = \frac{e^{-it E_\pi}}{2 \sqrt{\pi t |d_\alpha|}}e^{-i\, \rm{sgn}(d_\alpha) \frac{\pi}{4}}\,.
\end{align}
We again use integration by parts to obtain the second integral which constitutes the correction term $I_3^{\rm(corr)}$. Here the derivative is always finite in the interval $x \in [\delta, \infty)$, namely $\frac{\mathrm{d}}{\mathrm{d}x}(d_\alpha x^2)=2d_\alpha x \neq 0$ and can therefore be treated by the same integration by parts expansion used for $I_2$ as shown in Eq.~\eqref{eq:I2}. One will thus obtain,
\begin{align}
    I_3^{\rm(corr)}
    &= - \frac{e^{-it (E_\pi +d_\alpha x^2)}}{2it\pi d_\alpha x}\Bigg\vert_{\delta}^{\infty}
    + \frac{2d_\alpha\,e^{-it (E_\pi +d_\alpha x^2)}}{ (it)^2\pi (2d_\alpha x)^3}\Bigg\vert_{\delta}^{\infty}
    +\ldots =
    \frac{e^{-it(E_\pi+d_\alpha\delta^2)}}{2it\pi d_\alpha\delta}
    -\frac{e^{-it(E_\pi+d_\alpha\delta^2)}}{(it)^2\,4\pi d_\alpha^2\delta^3}
    +\ldots \,,
\end{align} 
which yields
\begin{equation}
\label{eq:I3}
    I_3 = \frac{e^{-it E_\pi}}{2 \sqrt{\pi t |d_\alpha|}}e^{-i\, \rm{sgn}(d_\alpha) \frac{\pi}{4}} -   \frac{e^{-it(E_\pi+d_\alpha\delta^2)}}{2it\pi d_\alpha\delta}
    +\frac{e^{-it(E_\pi+d_\alpha\delta^2)}}{(it)^2\,4\pi d_\alpha^2\delta^3}
    +\ldots.
\end{equation}
On the other hand, the local expansion near $k=\pi- \delta$, one has
\begin{align}
    E_k^\prime(\alpha)\big|_{k=\pi-\delta}&=-2d_\alpha\delta, \\
    E_k^{\prime \prime}(\alpha)\big|_{k=\pi-\delta}&=2d_\alpha.
\end{align}
Substituting these expressions into Eq.~\eqref{eq:I2}, the upper-endpoint contribution of $I_2$ at $k=\pi-\delta$ reads
\begin{align}
    \left.I_2\right|_{k=\pi-\delta}
    =
    \frac{e^{-it(E_\pi+d_\alpha\delta^2)}}{2it\pi d_\alpha\delta}
    -\frac{e^{-it(E_\pi+d_\alpha\delta^2)}}{(it)^2\,4\pi d_\alpha^2\delta^3}
    +\ldots,
\end{align}
that exactly cancels the endpoint contribution at $\pi-\delta$ of $I_3$ in Eq.~\eqref{eq:I3}.
Therefore, combining Eqs.~\eqref{eq:I1}, \eqref{eq:I2} and \eqref{eq:I3}, one obtains the Loschmidt amplitude from Eq.~\eqref{eq:Ln_detailed} as
\begin{align}\label{eq:ln_final}
    \mathcal{L}_n &= a_0\, e^{-in \tau E_0} (n)^{-1/p}  + a_{\pi} \, e^{-in\tau E_\pi} (n)^{-1/2}, 
\end{align}
with 
\begin{equation}
a_0 =\frac{1}{\pi p}(i c_{\alpha} \tau)^{-1/p} \,\Gamma\left(\frac{1}{p}\right), \quad \mbox{and} \quad a_{\pi}=\frac{1}{2\sqrt{\pi |d_{\alpha}| \tau}}e^{-i\mathrm{sgn}(d_{\alpha})\frac{\pi}{4}}.    
\label{eq:ln_final_coefficients}
\end{equation}
Equations \eqref{eq:ln_final} and \eqref{eq:ln_final_coefficients} have the form of Eq.~(9) of the main text. Modulus squared of this asymptotic formula is compared with exact numerical integration of the Loschmidt return probability, given by the modulus squared of \eqref{eq:ln_alphag1}, in Fig.~\ref{fig:figS2}(a) and (b) finding excellent agreement.

\subsection{Case for $\alpha = 2$}
\label{app:IIIE}
The expansion for $k^* = 0$ is still valid, but to obtain the value of $c_\alpha$ one needs to take the limiting values from the left and the right side. The local expansion follows $E_k(\alpha) = E_0 + c_\alpha |k|^{\alpha -1 }$, with $c_\alpha  = -2 \gamma \frac{\Gamma(1-\alpha) \cos(\frac{\pi}{2}(\alpha -1))}{\zeta(\alpha)}$. The factor $\Gamma(-1)$ has a pole at $-1$ and $|\Gamma(-1)| \to \infty$, whereas $\cos(\pi/2) = 0$. Taking the limiting value from the left and right side ($\alpha = 2 \pm \epsilon$), one can use the Euler reflection formula 
\begin{align}\label{eq:euler_ref}
    \Gamma(1-z)\Gamma(z) = \frac{\pi}{\sin(\pi z)} \,,
\end{align}
which is valid for $z \notin \mathbb{Z}$. One can now simplify $c_{2 \pm \epsilon}$ as
\begin{align}
    c_{2\pm \epsilon}  = -\gamma \frac{\pi}{\zeta(2 \pm \epsilon) \Gamma(2 \pm \epsilon) \cos(\frac{\pi }{2}(2 \pm \epsilon))},
\end{align}
which in the limit $\epsilon \to 0$ gives $c_2 =  \frac{6 \gamma}{\pi}$. Thus, one admits the local expansion 
\begin{align}
    E_k(2) = -2 \gamma + \frac{6 \gamma}{\pi}|k|\,.
\end{align}
Following the calculation in Subsec.~\ref{app:IIID}, one gets the Loschmidt amplitude as
\begin{align}\label{eq:ln_final_2}
    \mathcal{L}_n &= \frac{e^{-in \tau E_0}}{\pi} (i c_2 n \tau)^{-1} +  \frac{e^{-i n \tau E_\pi}}{2 \sqrt{\pi n \tau |d_2|}}e^{i \frac{\pi}{4}}= a_0 \, e^{-in \tau E_0} (n)^{-1}  + a_{\pi} \, e^{-in\tau E_\pi} (n)^{-1/2},
\end{align}
with $d_2 = - 3 \gamma/\pi^2$ and $E_\pi = \gamma$, $a_0 =(i c_2 \tau )^{-1}/\pi$ and $a_{\pi}=e^{i\pi/4}/(2 \sqrt{\pi |d_2| \tau})$. This result is again in the form of Eq.~(9) of the main text. Comparison between the Loschmidt return probability (modulus squared of Eq.~\eqref{eq:ln_final_2}) and numerical evaluation of the Loschmidt amplitude is reported in Fig.~\ref{fig:figS2}(c).


\subsection{Case for  $2<\alpha < 3$}
\label{app:IIIF}
For $2<\alpha<3$, $E_k(\alpha)$ has two stationary points at $k^*=0$ and $k^*=\pi$. $k^*=\pi$ is a regular quadratic stationary point, while $k^*=0$ is a nonanalytic stationary point, since it follows that the stationary point condition $E_k^\prime(\alpha)\vert_{k=0}=0$ is satisfied, but the second derivative
\begin{align}
    E_k^{\prime \prime}(\alpha)\sim p(p-1)c_\alpha k^{p-2}\,,
\end{align}
diverges as $k\to0$. Therefore, the contribution from $k^*=0$ cannot be treated by the standard Gaussian stationary-phase approximation and must be analyzed as in the case $1<\alpha<2$. Since the local expansion around $k^*=0, \pi$ remains of the same form as in Sec.~\ref{app:IIID}, one gets the same result for Loschmidt amplitude as Eq.~\eqref{eq:ln_final}, which is plotted in Fig.~\ref{fig:figS2}(d).


\subsection{Case for $\alpha = 3$}
\label{app:IIIG}
The local expansion at $k^* = 0$ follows 
\begin{align}
    E_k(3) = -2 \gamma \left( 1 + \frac{2 \log k - 3}{4 \zeta(3)} k^2\right)\,.
\end{align}
Using the leading order contribution, one gets for $k^* = 0$
\begin{align}\label{eq:integral_3}
    \mathcal{L}_n^{(0)} = \frac{e^{-it E_0}}{\pi} \int_0^\epsilon \mathrm{d}k e^{itc_3 k^2 \log k}\,,
\end{align}
where $c_3 = \gamma/\zeta(3)$ and $t = n\tau$. We now introduce a transformation $s = k^2 \log k$, one has $s<0$ for $0<k<\epsilon$. Setting $u = \log k$, one find $2s = 2ue^{2u}$ whose solutions are given by the Lambert-$W$ function~\cite{Defenu2019}. Since the large-$t$ asymptotics is controlled by the neighborhood of $k=0$ ($s<0$), we use the lower branch of the Lambert function.  For $- \frac{1}{2e} <  \epsilon^2 \log \epsilon < 0 $, one can invert the equation as $k = \exp(\frac{1}{2} W_{-1}(2s))$ and the derivative follows 
\begin{equation}
2k\, \mathrm{d}k = \frac{2 \, \mathrm{d}s}{1 + W_{-1}(2s)}, \quad \mbox{which gives} \quad \mathrm{d}k = \frac{\, \mathrm{d}s}{\exp(\frac{1}{2} W_{-1}(2s))[1 + W_{-1}(2s)]}. 
\end{equation}
Note that $\epsilon^2 \log \epsilon  < 0$, so we use $r = -s$. Substituting in the integral \eqref{eq:integral_3}, one gets
\begin{align}
      \mathcal{L}_n^{(0)} = \frac{1}{\pi} \int_0^{R} \mathrm{d}r\, \frac{  e^{-it E_0}\,e^{-itc_3r}}{\exp(\frac{1}{2} W_{-1}(-2r))[-1 - W_{-1}(-2r)]}\,,
\end{align}
where we defined $R = - \epsilon^2 \log \epsilon$. In the limit $r\to 0^+$, the function $W_{-1}(-2r)$ obeys the asymptotic expansion~\cite{Corless1996}
\begin{align}
    W_{-1}(-2r) = \log(2r) - \log( - \log (2r)) + \ldots\,.
\end{align}
By definition, $W_{-1}(-2r)e^{W_{-1}(-2r)} = -2r$, which yields $e^{\frac{1}{2}W_{-1}(-2r)} = (\frac{-2r}{W_{-1}(-2r)})^{\frac{1}{2}} \approx \sqrt{\frac{2r}{|\log 2r|}}$ and $-1 - W_{-1}(-2r) \approx |\log (2r)|$ in the leading order. Therefore, our integral can be approximated as
\begin{align}
    \mathcal{L}_n^{(0)} \approx \frac{e^{-it E_0}}{\sqrt{2}\pi} \int_0^{R} \mathrm{d}r\, \frac{r^{-1/2}e^{-itc_3r}}{\sqrt{-\log(2r)}}\,.
\end{align}
To proceed further, it is convenient to introduce the limit representation of logarithm $\log a = \lim_{h \to 0} (a^h - 1)/h$. This follows from the definition of derivative of $a^x$ at $x = 0$, i.e., $\frac{\mathrm{d}}{\rm{d}x} a^x = a^x \log a$. Since $a^0 = 1$, the limit representation gives the desired form of $\log a$~\cite{Defenu2019}. For our problem, we then represent
\begin{align}\label{eq:log_limit}
    -\log 2r = \lim_{h \to 0} \frac{1 - (2r)^h }{h}\,,
\end{align}
which in turn leads to
\begin{align}
    \frac{1}{\sqrt{-\log 2r}} = \lim_{h \to 0} \sqrt{h} \sum_{m = 0}^{\infty}C_m (2r)^{hm}\,,
\end{align}
where $C_m = \frac{2m !}{4^m (m!)^2}$. Plugging it into the integral one obtains
\begin{align}
    \mathcal{L}_n^{(0)} \approx\frac{e^{-in\tau E_0}}{\sqrt{2}\pi} \lim_{h \to 0} \sqrt{h}  \sum_{m = 0}^{\infty}C_m 2^{hm} \int_0^{R} \mathrm{d}r\, r^{hm-1/2}e^{-itc_3r}\,.
\end{align}
The remaining integral can be cast in terms of a lower incomplete Gamma function $\gamma(s,z)$:
\begin{align}
    \int_0^R \mathrm{d}r\, r^{hm-1/2}e^{-itc_3r} = (itc_3)^{-hm-1/2}\gamma\left(hm+\frac{1}{2},itc_3R\right)\,.
\end{align}
For $tc_3R \gg 1$, the incomplete Gamma function approaches the complete Gamma function (See also Sec.~\ref{app:IIID} for similar asymptotic analysis) and one obtains
\begin{align}
    \mathcal{L}_n^{(0)} &\approx \frac{e^{-it E_0}}{\sqrt{2 \pi}}(itc_3)^{-1/2}  \lim_{h \to 0} \sqrt{h}  \sum_{m = 0}^{\infty}C_m \left(\frac{2}{itc_3}\right)^{hm }\nonumber \\
    &\approx \frac{e^{-it E_0}}{\sqrt{2 \pi}}(itc_3)^{-1/2} \frac{1}{\sqrt{\log(itc_3/2)}} \approx \frac{e^{-it E_0}}{\sqrt{2 \pi c_3 t \log t}}e^{-i \pi/4}\,,
\label{eq:L_alpha_3_k0}
\end{align}
where we have used the limit $h \to 0$, to safely approximate $\Gamma(hm +1/2) \approx \Gamma(1/2) = \sqrt{\pi}$ (See also Appendix C of \cite{Defenu2019}) and approximated in the leading order $\log(itc_3/2) \approx \log t$. 

\begin{figure*}[!htpb]
\centering
\includegraphics[width=0.75\textwidth]{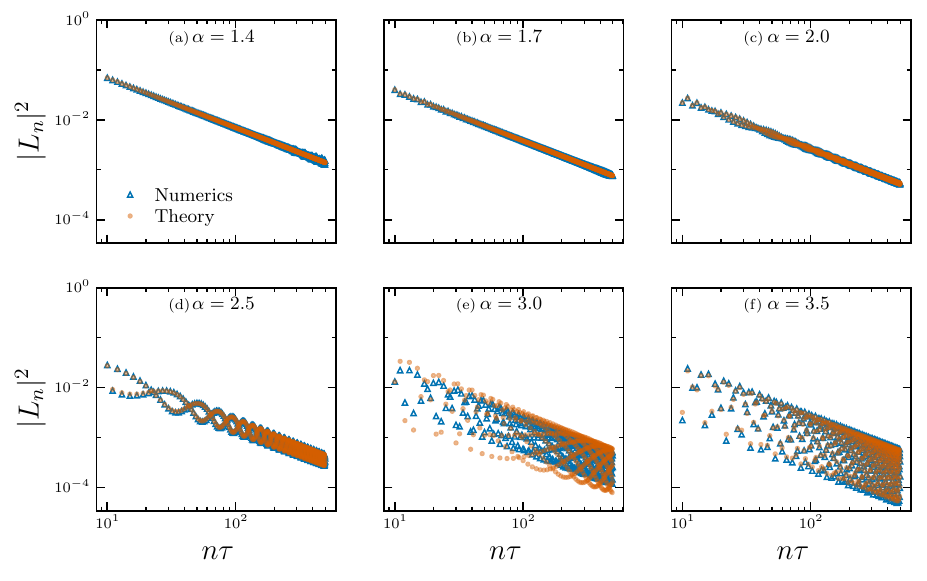}
\caption{\textit{Long-time asymptotics of the Loschmidt return probability.}
Log-log plot of the Loschmidt return probability $|\mathcal{L}_n|^2$ as a function of the stroboscopic time $n\tau$ for
(a) $\alpha=1.4$,
(b) $\alpha=1.7$,
(c) $\alpha=2.0$,
(d) $\alpha=2.5$,
(e) $\alpha=3.0$, and
(f) $\alpha=3.5$.
Blue triangles denote the numeric integration of the integral in Eq.~\eqref{eq:ln_alphag1}, while orange circles show the corresponding long-time asymptotic expressions derived in Sec.~III. Here $\gamma=1$ and $\tau=1$ and the numeric integration method used is Simpson integration with $\mathrm{d}k = 0.001$.
}
\label{fig:figS2}
\end{figure*}

For $k^* = \pi$, the expansion follows
\begin{align}
    E_k(3) &= -2 \gamma \left( \frac{-3}{4} + \frac{\log 2}{2\zeta(3)} (k - \pi)^2\right) = E_\pi^3 + d_3(k - \pi)^2\,,
\end{align}
where $E_\pi^3 = 3\gamma/2$ and $d_3 = - \gamma \log 2/\zeta(3)$. This gives contribution similar to Eq.~\eqref{eq:I3} and together with Eq.~\eqref{eq:L_alpha_3_k0} the asymptotics the Loschmidt amplitude as
\begin{align}
      \mathcal{L}_n \approx \frac{e^{-in\tau E_0}}{\sqrt{2 \pi c_3 n \tau \log (n \tau)}}e^{-i \pi/4} + \frac{e^{-in\tau E^3_\pi}}{2 \sqrt{\pi n \tau |d_3|}}e^{i\frac{\pi}{4}}=a_0 \frac{e^{-in\tau E_0}}{\sqrt{n \log (n \tau)}}+a_{\pi} \frac{e^{-in\tau E_{\pi}^3}}{\sqrt{n}},
      \label{eq:ln_alpha3}
\end{align}
with 
\begin{equation}
a_0=\frac{e^{-i\pi/4}}{\sqrt{2\pi c_3 \tau }}, \quad a_{\pi}= \frac{e^{i\pi/4}}{2 \sqrt{\pi |d_3| \tau }}.
\label{eq:a_coeff_alpha_3}
\end{equation}
The mode $k=0$ thus receives a logarithmic correction on top of the long-time algebraic scaling. This result is compared with exact numerics in Fig.~\ref{fig:figS2}(e).

\subsection{Case for  $\alpha > 3$}
\label{app:IIIH}
For $\alpha > 3$, both $k^* = 0, \pi$ are quadratic stationary points. At $k^*= 0$, the local expansion follows
\begin{align}
    E_k(\alpha) \approx  - 2\gamma  + \gamma  \frac{\zeta(\alpha - 2)}{\zeta(\alpha)} k^2  = E_0 + f_\alpha\, k^2\,, 
\end{align}
with $f_{\alpha}=\gamma \zeta(\alpha-2)/\zeta(\alpha)$. This gives the contribution
\begin{align}
    \mathcal{L}_n^{(0)} = \frac{e^{-i n \tau E_0}}{2 \sqrt{\pi n \tau |f_\alpha|}}
      e^{-i\frac{\pi}{4}},
\end{align}
and together with the contribution from $k^* = \pi$ (See Eq.~\eqref{eq:expansion_pi}), one obtains
\begin{align}\label{eq:ln_final_nn}
    \mathcal{L}_n = \frac{e^{-in \tau E_0}}{2 \sqrt{\pi n \tau |f_\alpha|}}
      e^{-i\frac{\pi}{4}} + \frac{e^{-i n\tau E_\pi}}{2 \sqrt{\pi n \tau |d_\alpha|}}e^{i \frac{\pi}{4}}=a_0 \frac{e^{-in\tau E_0}}{\sqrt{n}}+a_{\pi}\frac{e^{-in\tau E_{\pi}}}{\sqrt{n}},
\end{align}
with $a_0 = e^{-i\pi/4}/(2 \sqrt{ \pi  |f_\alpha| \tau })$ and  $a_\pi = e^{i\pi/4}/(2 \sqrt{ \pi  |d_\alpha| \tau })$. In this case, one thus sees that short and long wavelength modes contribute with the same exponent and the same spectral dimension to the long-time asymptotics of the Loschmidt amplitude. The long-time properties of the long-range quantum walk are thus equivalent to those of the short-range one. This result is plotted in Fig.~\ref{fig:figS2}(f).


\section{Detailed Calculation of the First-detection probability $F_n$ }
\label{app:IV}

\subsection{Case for $\alpha \leq 1$}
\label{app:IVA}
\begin{figure*}[!htpb]
\centering
\includegraphics[width=0.75\textwidth]{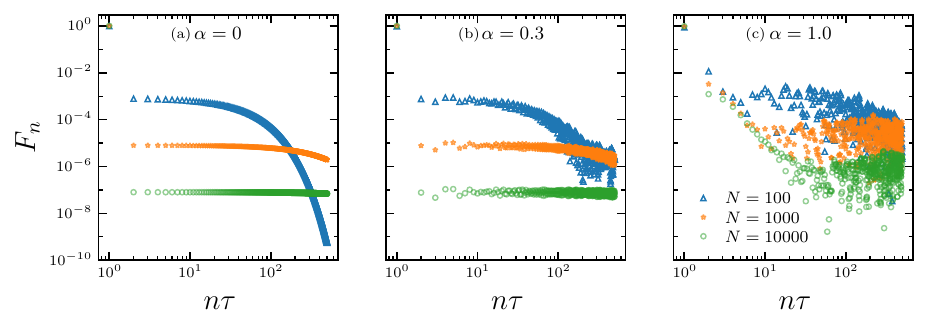}
\caption{\textit{First-detection probability in the recurrent regime.}
Log-log plot of the first-detection probability $F_n$ as a function of the stroboscopic time $n\tau$ for (a) $\alpha=0$,
(b) $\alpha=0.3$, and
(c) $\alpha=1$,
for system sizes $N=100$, $1000$, and $10000$. The first-detection probabilities are numerically obtained according to the procedure explained in Sec.~\ref{app:numerics}. For $\alpha\leq1$, increasing $N$ progressively suppresses the probability of detection at $n>1$. The return probability concentrates on the first attempt $n=1$ consistently with the thermodynamic-limit result $F_n=\delta_{n,1}$. Here $\gamma=1$ and $\tau=1$.} 
\label{fig:figS3}
\end{figure*}

As shown in Secs.~\ref{app:IIIA}, \ref{app:IIIB}, and \ref{app:IIIC}, the Loschmidt amplitude goes to $1$ for any $n$. Therefore the $z$-transform of the Loschmidt amplitude reduces to 
\begin{align}
      \tilde{\mathcal{L}}(z) &= \sum_{n=1}^\infty z^n  = \frac{z}{1- z}\,.
      \label{eq:lz_alpha0_approx}
\end{align}
Substituting Eq.~\eqref{eq:lz_alpha0_approx} into Eq.~\eqref{eq:phi_L_relation} gives $  \tilde{\phi}(z) = z$. Therefore, the first-detection amplitudes will be $\phi_n = \delta_{n,1}$, and hence, the corresponding first-detection probabilities will be $F_n = |\phi_n|^2 = \delta_{n,1}$. Physically, this result implies  detection is guaranteed at the first attempt; the particle remains essentially localized at the initial site and is detected with full certainty at the first-detection attempt. We show this result in Fig.~\ref{fig:figS3}, where we plot $F_n$ as a function of $n$ for three different values of $\alpha$ smaller than 1. In all the cases, we observe that upon increasing the system size $N$ the first-detection probability concentrates more and more on $n=1$, with negligible value for $n>1$. In the marginal case $\alpha=1$, this convergence is slower since it is logarithmic in the system size, as explained in Subsec.~\ref{app:IIIC}.

\subsection{Case for $1 < \alpha < 2$}
\label{app:IVB}
\begin{figure}[!htpb]
    \centering
        \includegraphics[scale =0.5]{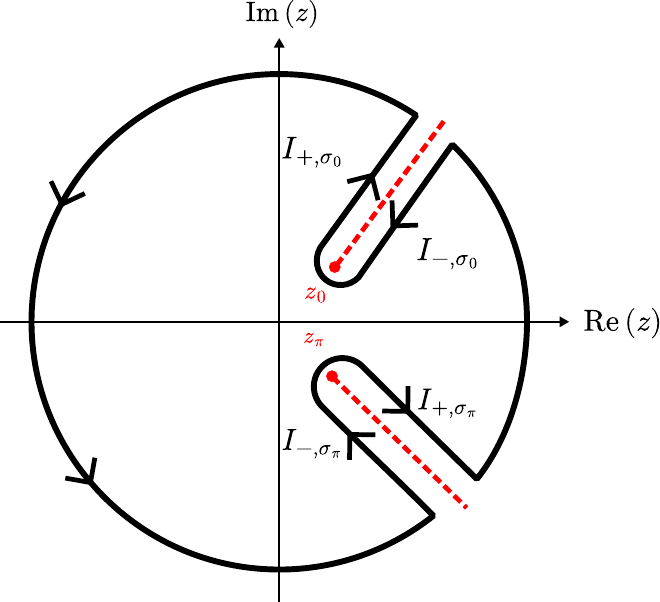}
        \caption{\textit{Integration contour.} Schematic representation of the contour used to evaluate the first-detection amplitude $\phi_n$ in the complex $z$ plane. The original contour enclosing the origin is deformed into two keyhole contours surrounding the branch cuts associated with the infrared and ultraviolet singularities of $\tilde{\mathcal L}(z)$. The branch points are located at $z_0=e^{iE_0\tau}$ and $z_\pi=e^{iE_\pi\tau}$, with the corresponding cuts parametrized as $z=r e^{iE_0\tau}$ and $z=r e^{iE_\pi\tau}$ for $r\geq1$. The contributions above and below each branch cut determine the long-time asymptotics of $\phi_n$.}
\label{fig:contour}
\end{figure}

 Using the asymptotic expansion of Loschmidt amplitudes from Eqs.~\eqref{eq:ln_final} and \eqref{eq:ln_final_coefficients}, one finds the $z$-transform of Loschmidt amplitude as
\begin{align}\label{eq:lz_simplified_1_alpha_3}
    \tilde{\mathcal{L}}(z) \approx a_0 \sum_{n=1}^\infty \frac{(ze^{-i  E_0 \tau})^n}{n^{\frac{1}{p}}} + a_\pi \sum_{n=1}^\infty \frac{(ze^{-i  E_\pi \tau})^n}{\sqrt{n}} \,,
\end{align}
where $a_0$ and $a_\pi$ are defined in Eq.~\eqref{eq:ln_final_coefficients}. Using the Polylogarithm function, one simplifies Eq.~\eqref{eq:lz_simplified_1_alpha_3} as
\begin{align} \label{eq:lz_polylog_1_alpha_3}
    \tilde{\mathcal{L}}(z) \approx a_0 \, \mathrm{Li}_{\frac{1}{p}}(ze^{-i E_0 \tau}) + a_\pi \, \mathrm{Li}_{\frac{1}{2}}(ze^{-i  E_\pi \tau})\,.
\end{align}
This result coincides with Eq.~(10) of the main text. For generic non-resonant $\tau$, this expression presents two branch-cuts along $z_0 = r \exp(iE_0\tau )$ and $z_\pi = r \exp(iE_\pi\tau ) $, and $r \geq 1$, due to the branch-cut singularity of $\mbox{Li}(z)$ along $z\in [1,\infty)$. We take henceforth the principal value $\mbox{Arg}(z)$ of the argument of a complex number $z$ in the interval $\mbox{Arg}(z)\in [-\pi,\pi)$. The first-detection amplitude $\phi_n$ is obtained by solving the contour integral in Eq.~\eqref{eq:inv_z_phi}, where the generating function $\tilde{\phi}(z)$ is given by Eq.~\eqref{eq:phi_L_relation}. When calculating the complex contour integral, the integration contour is deformed in a double keyhole contour enclosing both branch cuts as shown in Fig.~\ref{fig:contour}. Taking the limit of infinite radius $R$ of the contour, while extending the length of the part of the contour which encloses the branch cut at $z\in[1,\infty)$, the only nonvanishing contributions to the integral are provided by the line integrals above and below the two branch cuts. 
Analogous to Refs.~\cite{Friedman2017QuantumProblem,Walter2025ThermodynamicSystems}, we introduce
\begin{align}\label{eq:phi_n_inv_z_transform}
    I_{\pm, \sigma} &= \frac{1}{2 \pi i} \int_{1}^{\infty} \mathrm{d}(e^{i\sigma \tau}r) \frac{\tilde{\phi}((r\pm i0^+)e^{i\sigma\tau})}{[(r\pm i0^+)e^{i\sigma\tau}]^{n+1}} \nonumber \\
    &= \frac{e^{-in\sigma \tau}}{2 \pi i} \int_{1}^{\infty} \mathrm{d}r\, e^{-(n+1)\mathrm{ln}\,r} \tilde{\phi}((r\pm i0^+)e^{i\sigma\tau}) \nonumber \\
    &=  \frac{e^{-in\sigma \tau}}{2 \pi i} \int_{0}^{\infty}  \mathrm{d}\nu\, e^{-n \nu} \, \tilde{\phi}((e^\nu\pm i0^+)e^{i\sigma\tau}) \,,
\end{align}
where $\sigma_{0,\pi} = E_{0,\pi}$, and where we introduced $\nu = \rm{ln}\, r$, which is analytic along the positive real axis. In our notation, $I_{\pm, \sigma}$, the subscript $\sigma$ indicates which branch cut the integral is referring to, while $\pm$ indicates whether the contour was taken outwards $(+)$ or inwards $(-)$. By the residue theorem, the integral which determines $\phi_n$ then equals
\begin{align}\label{eq:phi_n_integral}
    \phi_n &= I_{+\,,\sigma_0} - I_{-\,,\sigma_0} + I_{+\,,\sigma_\pi} - I_{-\,,\sigma_\pi} \,. \nonumber \\
    &=   \sum_{\sigma} \frac{e^{-in\sigma \tau}}{2 \pi i} \int_{0}^{\infty}  \mathrm{d}\nu\, e^{-n \nu} \, [\tilde{\phi}_+^\sigma - \tilde{\phi}_-^\sigma]\,,
\end{align}
where $\tilde{\phi}_\pm^\sigma  = \tilde{\phi}((e^\nu\pm i0^+)e^{i\sigma\tau})$. Also, we can now relate the discontinuity in $\tilde{\phi}$ with the discontinuity in $\tilde{\mathcal{L}}$. This can be done exploiting that
\begin{align}\label{eq:disc_phi_disc_L}
    \tilde{\phi}_+^\sigma - \tilde{\phi}_-^\sigma &= \frac{\tilde{\mathcal{L}}_+^\sigma}{1 + \tilde{\mathcal{L}}_+^\sigma} - \frac{\tilde{\mathcal{L}}_-^\sigma}{1 + \tilde{\mathcal{L}}_-^\sigma} \nonumber \\&= \frac{\tilde{\mathcal{L}}_+^\sigma - \tilde{\mathcal{L}}_-^\sigma}{(1 + \tilde{\mathcal{L}}_+^\sigma )(1 + \tilde{\mathcal{L}}_-^\sigma)}\,,
\end{align}
where $\tilde{\mathcal{L}}_\pm^\sigma = \tilde{\mathcal{L}}((e^\nu\pm i0^+)e^{i\sigma\tau})$.

For large $n$, the integrand is exponentially suppressed and hence the integral is dominated by the behavior around $\nu = 0$. In order to study the resulting leading behavior, we expand the polylogarithm around $\nu = 0$~\cite{NIST:DLMF}. The leading terms of the polylog expansions are as follows

\begin{subequations}
\begin{align}
\operatorname{Li}_{1/p}(e^\nu \pm i0^+)
&= \Gamma\!\left(1-\frac{1}{p}\right)
   e^{\left(\frac{1}{p}-1\right)\log(-\nu \mp i0^+)}
   + \zeta\!\left(\frac{1}{p}\right)+\cdots ,
\label{eq:expansion_0}
\\
\operatorname{Li}_{1/2}(e^\nu \pm i0^+)
&= \sqrt{\pi}\,
   e^{-\frac{1}{2}\log(-\nu \mp i0^+)}
   + \zeta(1/2)+\cdots.
\label{eq:expansion_pi_2}
\end{align}
\end{subequations}
Here $\ldots$ denotes the sub-leading terms from the analytic terms in $\nu$.  For $\nu>0$, we use the expression of the branch-cut discontinuity of the logarithm along the negative real axis $\rm{ln}\,( -\nu \mp i0^+) = \rm{ln}\,(|-\nu| )\mp\, i \pi =  \rm{ln}\,\nu\, \mp \,i \pi$. On the branch cut corresponding to $\sigma_0$ we have, from Eq.~\eqref{eq:lz_polylog_1_alpha_3}, 
\begin{align}
    \tilde{\mathcal{L}}_\pm^{\sigma_0} = a_0\,\mathrm{Li}_{\frac{1}{p}}(e^\nu\pm i0^+) + \text{(analytic)}\,,
\end{align}
and thus using Eq.~\eqref{eq:expansion_0}, one gets
\begin{align}\label{eq:lz_diff}
    \tilde{\mathcal{L}}_+^{\sigma_0} -\tilde{\mathcal{L}}_-^{\sigma_0}
    &\approx a_0 \Gamma \left( 1- \frac{1}{p}\right) \left( e^{ (\frac{1}{p} - 1) \mathrm{ln}\,(-\nu - i0^+)} -  e^{ (\frac{1}{p} - 1) \mathrm{ln}\,(-\nu + i0^+)} \right) \nonumber \\
    &=a_0  \Gamma \left( 1 -  \frac{1}{p}\right)\nu^{(\frac{1}{p} - 1)}\left(e^{-i\pi(\frac{1}{p} - 1)}-e^{i\pi(\frac{1}{p} - 1)}\right) \nonumber \\
    &= -2i \sin\left(\frac{\pi(1 - p)}{p}\right) a_0\, \Gamma \left( 1 - \frac{1}{p}\right)\nu^{(\frac{1}{p} - 1)}\,.
\end{align}
 For $1<\alpha<2$, $\frac{1}{p} - 1 = \frac{2 - \alpha }{\alpha -1} >0$, here the nonanalytic term goes to $0$ with power law $\nu^{\frac{2 - \alpha}{\alpha -1}} \to 0$, so $\mathcal{\tilde{L}}_\pm^{\sigma_0}$ tends to a finite limit as $\nu \to 0^+$. Thus, $(1 + \tilde{\mathcal{L}}_+^{\sigma_0} )(1 + \tilde{\mathcal{L}}_-^{\sigma_0} ) \approx \left(1 + a_0 \zeta(\frac{1}{p}))\right)^2 \equiv D_0^2$. Thus, Eq.~\eqref{eq:disc_phi_disc_L} yields
\begin{align}\label{eq:disc_phi_leading_1_alpha_2}
    \tilde{\phi}_+^{\sigma_0} -\tilde{\phi}_-^{\sigma_0} 
    \approx \frac{-2i \sin\left(\frac{\pi(1 - p)}{p}\right)\,a_0\, \Gamma \left( 1 - \frac{1}{p}\right)}{D_0^2}\nu^{(\frac{1}{p} - 1)}\, .
\end{align}
Inserting Eq.~\eqref{eq:disc_phi_leading_1_alpha_2} into the keyhole-contour representation (See Eq.~\eqref{eq:phi_n_integral}), we obtain
\begin{align}
    I_{+,\sigma_0}-I_{-,\sigma_0}
    &\approx \frac{e^{-in\sigma_0\tau}}{2\pi i}\int_0^\infty d\nu\,e^{-n\nu}\,
   A_0\,\nu^{(\frac{1}{p} - 1)}\,,
\end{align}
where $A_0 = - \frac{2i \sin\left(\frac{\pi(1 - p)}{p}\right)\,a_0\, \Gamma \left( 1 - \frac{1}{p}\right)}{D_0^2}$.
Evaluating the integral, one obtains
\begin{align}
    \int_0^\infty d\nu\,\nu^{(\frac{1}{p} - 1)}e^{-n\nu}=\Gamma(1/p)\,n^{-\frac{1}{p}}.
\end{align}
One can further simplify using Euler reflection formula(See Eq.~\eqref{eq:euler_ref}) as follows
\begin{align}
    I_{+,\sigma_0}-I_{-,\sigma_0}
   &\approx A_0 \frac{e^{-in\sigma_0\tau}}{2\pi i} \Gamma(1/p) \,n^{-\frac{1}{p}} = \frac{a_0\, e^{-in\sigma_0\tau} \,  n^{-1/p}}{D_0^2}.
    \label{eq:I_sigma0_asymptotic}
\end{align}
An analogous calculation for the second branch cut ($\sigma_\pi$) yields
\begin{align}
    I_{+,\sigma_\pi}-I_{-,\sigma_\pi}
    \approx \frac{e^{-in\sigma_\pi\tau}}{2\pi a_\pi}\,n^{-3/2}.
    \label{eq:I_sigmap_asymptotic}
\end{align}

Combining Eq.~\eqref{eq:I_sigma0_asymptotic} and Eq.~\eqref{eq:I_sigmap_asymptotic}, one obtains the first-detection amplitude as
\begin{align}
    \phi_n \approx  \frac{a_0 e^{-in\sigma_0\tau}}{D_0^2} n^{-1/p} + \frac{e^{-in\sigma_\pi\tau}}{2\pi a_\pi}\,n^{-3/2}\,,
\end{align}
which gives the first-detection probability as
\begin{align}
     F_n \approx \left\vert  \frac{a_0 e^{-in\sigma_0\tau}}{D_0^2} n^{-1/p} + \frac{e^{-in\sigma_\pi\tau}}{2\pi a_\pi}\,n^{-3/2} \right\vert^2\,. 
\label{eq:final_f_n_5/3}
\end{align}
Thus we arrive at a conclusion that the tails of $\phi_n \sim n^{-3/2}$ for $ 1<\alpha \leq 5/3$ and $\phi_n \sim n^{- \frac{1}{\alpha -1}}$ for $5/3 \leq \alpha < 2$. Thus the first-detection probability decays like  $F_n \sim n^{-3}$ for $ 1<\alpha \leq 5/3$ and $F_n \sim n^{- \frac{2}{\alpha -1}}$ for $5/3 \leq \alpha < 2$. The result \eqref{eq:final_f_n_5/3} is compared with numerical evaluation of $F_n$ in Fig.~3(c) of the main text. In Figs.~\ref{fig:figS5}(a) and (b) we further assess the dependence of $F_n$ on the system size $N$. It is clear that upon increasing the system size, the agreement with \eqref{eq:final_f_n_5/3} improves. For small system size, instead, finite size effects lead to a deviation from the thermodynamic limit results.


\subsection{Case for $\alpha = 2$}
\label{app:IVC}

Following the same procedure as in Sec.~\ref{app:IVB}, one obtains using Eq.~\eqref{eq:ln_final_2} the $z$-transform of Loschmidt amplitude as

\begin{align} \label{eq:lz_polylog_2}
    \tilde{\mathcal{L}}(z) \approx a_0 \, \mathrm{Li}_1(ze^{-i E_0 \tau}) + a_\pi \, \mathrm{Li}_{\frac{1}{2}}(ze^{-iE_\pi \tau})\,,
\end{align}
where $a_0$ and $a_\pi$ are defined after Eq.~\eqref{eq:ln_final_2}. Using the same procedure as in Sec.~\ref{app:IVB}, we first write the leading terms of the polylog expansions~\cite{NIST:DLMF}
\begin{align}
&\mathrm{Li}_{1}(e^\nu \pm i0^+)
= - \log(1 - (e^\nu \pm i0^+)) \nonumber \\&= - \log(1 - e^\nu \mp i0^+) = - \log(e^\nu - 1) \pm i \pi,
\end{align}
and the expansion for the other polylogarithm follows Eq.~\eqref{eq:expansion_pi_2}. On the branch cut corresponding to $\sigma_0$, we have approximated
\begin{align}
      \tilde{\mathcal{L}}_+^{\sigma_0} -\tilde{\mathcal{L}}_-^{\sigma_0} &\approx 2 \pi i a_0\,, \\
      (1 + \tilde{\mathcal{L}}_+^{\sigma_0} )(1 + \tilde{\mathcal{L}}_-^{\sigma_0} ) &\approx [-a_0 \log(e^\nu - 1)]^2\,.
\end{align}
Using Eq.~\eqref{eq:disc_phi_disc_L} will give in the leading orders
\begin{align}\label{eq:disc_phi_leading_2}
\tilde{\phi}_+^{\sigma_0} -\tilde{\phi}_-^{\sigma_0} 
\approx \frac{2 \pi i}{a_0 [-\log(e^\nu - 1)]^2}
\end{align}
Inserting Eq.~\eqref{eq:disc_phi_leading_2} into the keyhole-contour representation (See Eq.~\eqref{eq:phi_n_integral}), we obtain
\begin{align}
    I_{+,\sigma_0}-I_{-,\sigma_0}
    &\approx \frac{e^{-in\sigma_0\tau}}{a_0}\int_0^\infty d\nu\,\frac{e^{-n\nu}}{ [-\log(e^\nu - 1)]^2}\approx\frac{e^{-in\sigma_0\tau}}{a_0}\int_0^\infty d\nu\,\frac{e^{-n\nu}}{ [-\log(\nu)]^2}\,.
\end{align}
In order to proceed further we use the limit representation of the logarithm as given in Eq.~\eqref{eq:log_limit}, which in turn leads to
\begin{align}
    \frac{1}{[-\log \nu]^2} = \lim_{h \to 0} h^2\sum_{m = 0}^\infty (m +1)\nu^{hm}\,.
\end{align}
Plugging this back into the integral, one gets
\begin{align}
    I_{+,\sigma_0}-I_{-,\sigma_0}
   &\approx\frac{e^{-in\sigma_0\tau}}{a_0} \lim_{h \to 0} h^2 \sum_{m = 0}^\infty (m +1)\int_0^\infty d\nu\,e^{-n\nu} \nu^{hm} = \frac{e^{-in\sigma_0\tau}}{a_0} \lim_{h \to 0} h^2 \sum_{m = 0}^\infty (m +1) n^{-hm -1}\Gamma(hm + 1) \nonumber \\
   &\approx \frac{e^{-in\sigma_0\tau}}{a_0\, n} \lim_{h \to 0} h^2 \sum_{m = 0}^\infty (m +1) n^{-hm}\approx \frac{e^{-in\sigma_0\tau}}{a_0\,n} \lim_{h \to 0} \left( \frac{h}{1 - n^{-h}} \right)^2 \approx \frac{e^{-in\sigma_0\tau}}{a_0\,n} \frac{1}{[\log n]^2} \label{eq:I_sigma0_asymptotic2}\,,
\end{align}
where we have approximated $\Gamma(1 + hm) \approx 1$ (See also Sec.~\ref{app:IIIG} and Appendix C of \cite{Defenu2019}). An analogous calculation for the second branch cut ($\sigma_\pi$) yields
\begin{align}
    I_{+,\sigma_\pi}-I_{-,\sigma_\pi}
    \approx \frac{e^{-in\sigma_\pi\tau}}{2\pi a_\pi}\,n^{-3/2}.
    \label{eq:I_sigmap_asymptotic2}
\end{align}
\noindent 

\begin{figure*}[!htpb]
\centering
\includegraphics[width=0.9\textwidth]{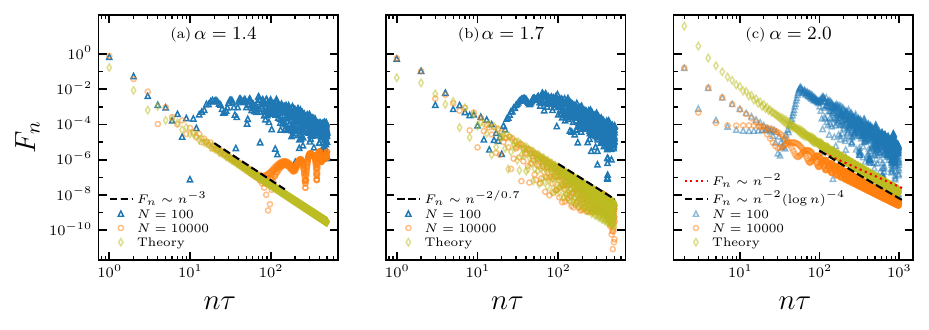}
\caption{\textit{First-detection probability in the transient regime for $1<\alpha\leq2$.}
Log-log plot of the first-detection probability $F_n$ as a function of the stroboscopic time $n\tau$ for (a) $\alpha=1.4$,
(b) $\alpha=1.7$, and
(c) $\alpha=2$.
In panels (a) and (b), different symbols show exact numerical results for $N=100$ and $10000$. Analytical asymptotic predictions are shown by the diamond markers from Eq.~\eqref{eq:final_f_n_5/3}. The black dashed lines are obtained from numerical fitting with the power-law model indicated in the corresponding panel. In panel (c), theoretical predictions from Eq.~\eqref{eq:final_F_alpha_2} are shown with diamond markers. The black-dashed line is obtained from numerical fitting including logarithmic corrections $F_n\sim n^{-2} (\log n)^{-4}$. The red-dotted line is a numerical fitting with the algebraic asymptotic $n^{-2}$ and it is reported for comparison. Here $\gamma=1$ and $\tau=1$.} 
\label{fig:figS5}
\end{figure*}

Combining Eq.~\eqref{eq:I_sigma0_asymptotic2} and Eq.~\eqref{eq:I_sigmap_asymptotic2}, one obtains the first-detection amplitude as
\begin{align}
    \phi_n \approx \frac{e^{-in\sigma_0\tau}}{a_0} \frac{1}{n[\log n]^2} + \frac{e^{-in\sigma_\pi\tau}}{2\pi a_\pi}\,n^{-3/2}\,,
\end{align}
which gives the first-detection probability as
\begin{align}
     F_n \approx \left\vert  \frac{e^{-in\sigma_0\tau}}{a_0} \frac{1}{n[\log n]^2} + \frac{e^{-in\sigma_\pi\tau}}{2\pi a_\pi}\,n^{-3/2} \right\vert^2\,. 
\label{eq:final_F_alpha_2}
\end{align}
For large $n$, the first-detection amplitude decays like $\phi_n \sim 1/(n [\log n]^2)$ which leads to the first-detection probability to decay like $F_n \sim 1/(n^2 [\log n]^4)$. This result is shown in Fig.~\ref{fig:figS5}(c). The asymptotics matches the numerical data for large system sizes $N$ and long times. We note that in this case longer times than in Figs.~\ref{fig:figS5}(a) and (b) are needed in order for the numerical data. This is caused by a slowly decaying logarithmic correction in Eq.~\eqref{eq:final_F_alpha_2}.

\subsection{Case for $2 < \alpha < 3$}
\label{app:IVD}

 For $2<\alpha<3$, $\frac{1}{p} - 1 = \frac{2- \alpha}{\alpha -1} < 0$, hence the nonanalytic term diverges with $\nu^{\frac{2- \alpha}{\alpha -1}}$, so the leading contribution to the denominator is given by the product $\tilde{\mathcal L}_+^{\sigma_0}\tilde{\mathcal L}_-^{\sigma_0} \approx a_0^2 \Gamma^2(1 - \frac{1}{p})\nu^{2(\frac{1}{p} - 1)}$, where $a_0$ is defined in Eq.~\eqref{eq:ln_final_coefficients}. Thus, Eq.~\eqref{eq:disc_phi_disc_L} and Eq.~\eqref{eq:lz_diff} yield

\begin{align}\label{eq:disc_phi_leading_2_alpha_3}
    \tilde{\phi}_+^{\sigma_0} -\tilde{\phi}_-^{\sigma_0} 
    \approx- \frac{2 i  \sin\left(\frac{\pi(1 - p)}{p}\right) }{a_0 \Gamma(1 - \frac{1}{p})} \nu^{-(\frac{1}{p} - 1)}\,.
\end{align}
Inserting Eq.~\eqref{eq:disc_phi_leading_2_alpha_3} into the keyhole-contour representation (See Eq.~\eqref{eq:phi_n_integral}), we obtain
\begin{align}
    & I_{+,\sigma_0}-I_{-,\sigma_0} 
    \approx - \frac{2 i  \sin\left(\frac{\pi(1 - p)}{p}\right) }{a_0 \Gamma(1 - \frac{1}{p})}  \frac{e^{-in\sigma_0\tau}}{2\pi i}\int_0^\infty d\nu\,e^{-n\nu}\,
   \,\nu^{-(\frac{1}{p} - 1)}\,.
\end{align}
Solving the integral and simplifying using the property of Gamma function $\Gamma(1 + z) = z\Gamma(z)$, one obtains
\begin{align}
    I_{+,\sigma_0}-I_{-,\sigma_0}
    \approx \frac{(1 - \frac{1}{p})\sin(\frac{\pi}{p}) e^{-in\sigma_0\tau}}{\pi a_0} \,n^{-(2 - \frac{1}{p})}.
    \label{eq:I_sigma0_asymptotic_2}
\end{align}
An analogous calculation for the second branch cut ($\sigma_\pi$) yields
\begin{align}
    I_{+,\sigma_\pi}-I_{-,\sigma_\pi}
    \approx \frac{e^{-in\sigma_\pi\tau}}{2\pi a_\pi}\,n^{-3/2},
    \label{eq:I_sigmap_asymptotic_2}
\end{align}
where $a_\pi$ is defined in Eq.~\eqref{eq:ln_final_coefficients}. Combining Eq.~\eqref{eq:I_sigma0_asymptotic_2} and Eq.~\eqref{eq:I_sigmap_asymptotic_2}, one obtains the first-detection amplitude as
\begin{align}
    \phi_n \approx \frac{(1 - \frac{1}{p})\sin(\frac{\pi}{p}) e^{-in\sigma_0\tau}}{\pi a_0} \,n^{-(2 - \frac{1}{p})} + \frac{e^{-in\sigma_\pi\tau}}{2\pi a_\pi}\,n^{-3/2}\,,
\end{align}
which gives the first-detection probability as
\begin{align}
     F_n \approx \left\vert \frac{(1 - \frac{1}{p})\sin(\frac{\pi}{p}) e^{-in\sigma_0\tau}}{\pi a_0} \,n^{-(2 - \frac{1}{p})} + \frac{e^{-in\sigma_\pi\tau}}{2\pi a_\pi}\,n^{-3/2} \right\vert^2\,. 
\label{eq:fn_final_2p5}
\end{align}
For large $n$, the first-detection amplitude decays like $\phi_n \sim n^{-(2 - \frac{1}{\alpha -1})}$ which leads to the first-detection probability to decay like $F_n \sim n^{-(4 - \frac{2}{\alpha -1})}$. This result is compared in Fig.~\ref{fig:figS6}(a) with numerical data for different system sizes.

\subsection{Case for $\alpha = 3$}
\label{app:IVE}
Using Eq.~\eqref{eq:ln_alpha3}, one can write the $z$-transform of Loschmidt amplitude as
\begin{align}
\tilde{\mathcal{L}}(z) \simeq  a_0 \,\sum_{n\ge 2}\frac{\left(z e^{-iE_0\tau}\right)^{n}}{\sqrt{n \log n}}
  + a_\pi\,\mathrm{Li}_{1/2} \left(z e^{-iE_\pi\tau}\right)  ,
\end{align}
with  $a_{0,\pi}$ defined in Eq.~\eqref{eq:a_coeff_alpha_3}. 
We first define 
\begin{align}
    \mathcal{K}(z) \equiv \sum_{n\ge 2}\frac{z^{n}}{\sqrt{n \log n}}\,.
\end{align}
The integral in Eq.~\ref{eq:phi_n_inv_z_transform} for the $\sigma_0$ branch is now given by $\mathcal{K}(e^\nu \pm i 0^+)$ around $\nu = 0$ for large $n$. The first step is to express $\mathcal{K} (z)$ in terms of polylogarithm function. Using the limit representation of logarithm as in Eq.~\eqref{eq:log_limit}, one gets
\begin{align}
    \mathcal{K}(z) = \sum_{n\geq 2} \frac{z^n}{n^{1/2}}\frac{1}{\sqrt{\log n}} = \lim_{h \to 0}   \sqrt{h}  \sum_{m = 0}^{\infty} \sum_{n\geq 2} \frac{z^n}{n^{1/2}} C_m n^{-hm} = \lim_{h \to 0}   \sqrt{h}  \sum_{m = 0}^{\infty} C_m\, \left( \mathrm{Li}_{1/2 + hm}(z) - z\right)\,,
\end{align}
with $C_m = \frac{2m !}{4^m (m!)^2} $. The non-analytic contribution comes from the polylogarithm and hence the leading terms of the expansion are as follows
\begin{align}
    \mathcal{K}(e^\nu \pm i0^+) &=  \lim_{h \to 0}   \sqrt{h}  \sum_{m = 0}^{\infty} C_m \Gamma\left(\frac{1}{2} - hm\right) e^{(hm - 1/2) \log (- \nu \mp i0^+) } + \ldots \nonumber \\
    &= \Gamma(1/2)e^{- \frac{1}{2}\log (- \nu \mp i0^+) } \lim_{h \to 0}   \sqrt{h}  \sum_{m = 0}^{\infty} C_m (- \nu)^{hm} \approx \sqrt{\pi} e^{- \frac{1}{2}\log (- \nu \mp i0^+) } \frac{1}{\sqrt{\log (1/\nu)} }
\nonumber \\ &= \pm \frac{i \sqrt{\pi} }{\sqrt{\nu \log (1/\nu)}}\,.
\end{align}
We thus get
\begin{align}\label{eq:K_alpha3}
    \mathcal{K}(e^\nu + i0^+) -  \mathcal{K}(e^\nu - i0^+) &\approx 2i \frac{\sqrt{\pi} }{\sqrt{\nu \log (1/\nu)}}\,, \\
   \mathcal{K}(e^\nu + i0^+)   \mathcal{K}(e^\nu - i0^+) &= \frac{\pi}{\nu \log (1/\nu)}\,.
\end{align}
It follows then 
\begin{align}\label{eq:disc_phi_leading_alpha_3}
    \tilde{\phi}_+^{\sigma_0} -\tilde{\phi}_-^{\sigma_0} \approx  \frac{2 i }{a_0 \sqrt{\pi}} \sqrt{\nu \log (1/\nu)}.
\end{align}
Inserting Eq.~\eqref{eq:disc_phi_leading_alpha_3} into the keyhole-contour representation(See Eq.~\eqref{eq:phi_n_integral}), we obtain
\begin{align}
    I_{+,\sigma_0}-I_{-,\sigma_0}  \approx  \frac{2 i }{a_0 \sqrt{\pi}} \frac{e^{-i n \sigma_0 \tau}}{2 \pi i} \int_{0}^\infty \mathrm{d}\nu \, e^{-n \nu }  \sqrt{\nu \log (1/\nu)}.
\end{align}
Performing a change of variable to $x = n \nu$, one gets
\begin{align}
    I_{+,\sigma_0}-I_{-,\sigma_0} &\approx  \frac{2 i }{a_0 \sqrt{\pi}} \frac{e^{-i n \sigma_0 \tau}}{2 \pi i} \frac{1}{n^{3/2}}\int_{0}^\infty \mathrm{d}x \, e^{-x }  \sqrt{x} \sqrt{\log n - \log x} \nonumber \\
    &\approx  \frac{2 i }{a_0 \sqrt{\pi}} \frac{e^{-i n \sigma_0 \tau}}{2 \pi i} \frac{1}{n^{3/2}}\int_{0}^\infty \mathrm{d}x \, e^{-x }  \sqrt{x} \sqrt{\log n} =  \frac{2 i }{a_0 \sqrt{\pi}} \frac{e^{-i n \sigma_0 \tau}}{2 \pi i}  \frac{\sqrt{\log n}}{n^{3/2}} \Gamma(3/2) = \frac{e^{-i n \sigma_0 \tau}}{2 \pi a_0}  \frac{\sqrt{\log n}}{n^{3/2}}.
    \label{eq:I_sigma0_asymptotic_3}
\end{align}
An analogous calculation for the second branch cut ($\sigma_\pi$) yields
\begin{align}
    I_{+,\sigma_\pi}-I_{-,\sigma_\pi}
    \approx \frac{e^{-in\sigma_\pi\tau}}{2\pi a_\pi}\,n^{-3/2}.
    \label{eq:I_sigmap_asymptotic_3}
\end{align}

Combining Eq.~\eqref{eq:I_sigma0_asymptotic_3} and Eq.~\eqref{eq:I_sigmap_asymptotic_3}, one obtains the first-detection amplitude as
\begin{align}
    \phi_n \approx  \frac{e^{-i n \sigma_0 \tau}}{2 \pi a_0}  \sqrt{\log n}\, n^{-3/2} + \frac{e^{-in\sigma_\pi\tau}}{2\pi a_\pi}\,n^{-3/2}\,,
\end{align}
which gives the first-detection probability as 
\begin{align}
    F_n= \left\vert  \frac{e^{-i n \sigma_0 \tau}}{2 \pi a_0}  \sqrt{\log n}\, n^{-3/2} + \frac{e^{-in\sigma_\pi\tau}}{2\pi a_\pi}\,n^{-3/2} \right\vert^2\,.
\label{eq:fn_final_3}
\end{align}
For large $n$, the first-detection probability decays as $F_n \sim \log n \,\, n^{-3}$. This result is shown in Fig.~\ref{fig:figS6}(b). It correctly reproduces the numerical data for large system sizes.

\subsection{Case for $\alpha >3$}
\label{app:IVF}

\begin{figure*}[!htpb]
\centering
\includegraphics[width=0.9\textwidth]{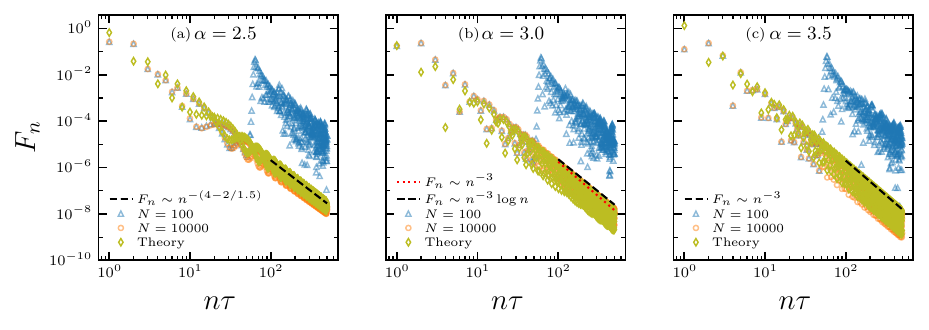}
\caption{\textit{First-detection probability in the transient regime for $\alpha>2$.}
Log-log plot of the first-detection probability $F_n$ as a function of the stroboscopic time $n\tau$ for (a) $\alpha=2.5$,
(b) $\alpha=3$, and
(c) $\alpha=3.5$. Different symbols show exact numerical results obtained for $N=100$ and $10000$. The analytical asymptotic prediction is shown by the diamond markers and it corresponds to Eq.~\eqref{eq:fn_final_2p5} for panel (a), Eq.~\eqref{eq:fn_final_3} for panel (b) and Eq.~\eqref{eq:fn_final_alpha_g3} for panel (c). Black dashed lines are obtained via fitting with the model indicated in the corresponding panel. In panel (b), the red-dotted line is also a fitting with the algebraic form $F_n \sim n^{-3}$ and it is reported for comparison with the actual asymptotics featuring logarithmic correction. Here $\gamma=1$ and $\tau=1$.} 
\label{fig:figS6}
\end{figure*}

For hopping exponent $\alpha>3$, the system is qualitatively similar to the case of nearest neighbor hopping. The latter was analyzed in detail in Ref.~\cite{Friedman2017QuantumProblem}. Using the asymptotic expansion of Loschmidt amplitudes from Eq.~\eqref{eq:ln_final_nn}, one finds the $z$-transform of the Loschmidt amplitude as
\begin{align} \label{eq:simplified_lz}
    \tilde{\mathcal{L}}(z) \approx a_0 \, \mathrm{Li}_{1/2}(ze^{-i  E_0 \tau}) + a_\pi \, \mathrm{Li}_{1/2}(ze^{-i E_\pi \tau})\,,
\end{align}
where $a_{0,\pi}$ are defined in Eq.~\eqref{eq:ln_final_nn}. Analogous to calculation in previous subsections, we obtain the first-detection amplitude using Eq.~\eqref{eq:phi_n_integral} as
\begin{align}
    \phi_n \approx \frac{1}{2\pi}
    \left(  \frac{e^{-in\sigma_0\tau}}{a_0} + \frac{e^{-in\sigma_\pi\tau}}{a_\pi} \right) n^{-3/2},
\end{align}
Thus the first-detection probability follows
\begin{align}
   F_n \approx\frac{1}{4\pi^2}
    \left|  \frac{e^{-in\sigma_0\tau}}{a_0} + \frac{e^{-in\sigma_\pi\tau}}{a_\pi} \right|^2 n^{-3}.
\label{eq:fn_final_alpha_g3}
\end{align}
This result is shown in Fig.~\ref{fig:figS6}(c) together with numerical data for different system sizes.

\section{Resonant probing times $\tau_{\rm{res}}$}
\begin{figure*}[!htpb]
\centering
\includegraphics[width=0.9\textwidth]{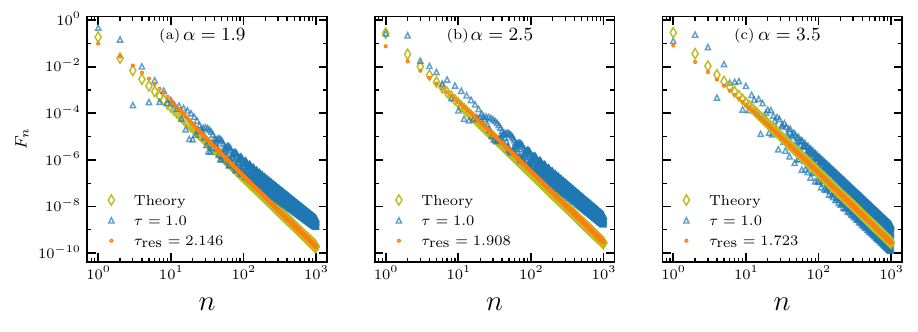}
\caption{\textit{First-detection probability at resonant probing times.} Log-log plot of the first-detection probability $F_n$ as a function of the measurement attempt $n$ for
(a) $\alpha=1.9$, (b) $\alpha=2.5$, and (c) $\alpha=3.5$. Blue triangles show the numerical results at the generic probing time $\tau=1$, while orange circles correspond to numerical results at the resonant probing time, $\tau_{\rm res}$ according to Eq.~\eqref{eq:resonant_tau}. The olive diamonds indicate the theory curve obtained in Eq.~\eqref{eq:res_alpha_l3} for (a) and (b) and Eq.~\eqref{eq:res_alpha_g3} for (c). The theory curves indicate the $n^{-3}$ asymptotic behavior. The oscillations arising from the interference between the two branch-point contributions at generic probing times are absent at resonant probing times.
Here $\gamma=1$ and $N = 10^6$.}
\label{fig:figS7}
\end{figure*}

So far, we have used that the two branch points corresponding to the infrared $k = 0$ and ultraviolet $k = \pi$ modes are distinct, but there exist special probing times $\tau_{\rm{res}}$ for which the two branch points coincide. We refer to these points as the resonant probing times, as done also in Refs.~\cite{Friedman2017QuantumProblem,yin2025restart,Walter2025ThermodynamicSystems}. For $\alpha>1$, the two branch points of the generating function $\tilde{\mathcal L}(z)$ are located at
$z_0=e^{iE_0\tau}, z_\pi=e^{i E_\pi\tau}$. A resonance occurs when $z_0=z_\pi$ or, equivalently,
\begin{align}
    \left(E_\pi-E_0\right) \tau_{\rm{res}} =2\pi m,
    \qquad m=1,2,\ldots .
    \label{eq:resonance_condition}
\end{align}
Using $E_0 =-2\gamma$, and $E_\pi=-2\gamma\left(2^{1-\alpha}-1\right)$, the resonant probing times are
\begin{align}
   \tau_{\rm{res}} =
    \frac{\pi m}
    {2\gamma\left(1-2^{-\alpha}\right)}\,.
    \label{eq:resonant_tau}
\end{align}
At $\tau_{\rm {res}}$, one finds the $z$-transform of Loschmidt amplitude as
\begin{align} \label{eq:res_Lz}
    \tilde{\mathcal{L}}(z) \approx a_0 \mathrm{Li}_{\frac{1}{p}}(ze^{-iE_0 \tau_{\rm{res}}}) +  a_\pi \mathrm{Li}_{\frac{1}{2}}(ze^{-iE_\pi \tau_{\rm{res}}})\,.
\end{align}
There will be a single branch cut at $z_{\rm{res}} = re^{i E_0 \tau_{\rm {res}}} = re^{i E_\pi \tau_{\rm {res}}}$~\cite{Friedman2017QuantumProblem,Walter2025ThermodynamicSystems}. We now determine which of the two contributions controls the singular behavior. As in Subsec.~\ref{app:IVB}, we introduce $\nu = \log r$, and hence on the contours ($\pm$), the values of the generating function is given by
\begin{align}
    \tilde{\mathcal{L}}_{\pm} \approx a_0 \mathrm{Li}_{\frac{1}{p}}(e^\nu \pm i0^+) + a_\pi \mathrm{Li}_{\frac{1}{2}}(e^\nu \pm i0^+) \approx a_0 \Gamma\left(1-\frac{1}{p}\right)
e^{\mp i\pi(\frac{1}{p}-1)} \nu^{\frac{1}{p} - 1}  \pm i a_\pi \sqrt{\pi} \nu^{-1/2} + \rm{analytic} \approx \pm i a_{\pi} \sqrt{\pi} \nu^{-1/2}\,.
\label{eq:singularity_resonance}
\end{align}
The reasoning is as follows; for $1< \alpha \leq 2$, $1/p - 1 \geq 0$ which implies $\nu^{1/p - 1} \to 0$ and thus the singular behavior is governed by $\nu^{-1/2}$. For $2< \alpha < 3$, $-1/2 < 1/p -1 < 0$, which implies $\nu^{-1/2} \gg \nu^{1/p -1}$. At $\alpha = 3$, the $k = 0$ contribution is logarithmically suppressed (See Sec.~\ref{app:IVE}). The singular behavior in Eq.~\eqref{eq:singularity_resonance} at $\nu=0$ is therefore always governed by the ultraviolet mode $k=\pi$. The first-detection probability asymptotics for $1<\alpha \leq 3$ and resonant $\tau$ values is accordingly given by the short-range result
\begin{align}\label{eq:res_alpha_l3}
    F_n \approx \frac{1}{4 \pi^2 \vert a_\pi\vert^2}\, n^{-3}\,.
\end{align}
However, for $\alpha> 3$, both $k = 0$ and $k = \pi$ have square root singularities, and hence the amplitudes need to be added, which gives 
\begin{align}\label{eq:res_alpha_g3}
    F_n \approx \frac{1}{4 \pi^2\vert a_0 +a_\pi \vert^2} n^{-3}\,.
\end{align}
Note that no oscillations are present in $F_n$ for resonant probing times because of a single branch cut and it decays like $n^{-3}$ for all $\alpha > 1$ with prefactor dependent on $\alpha$. This is shown in Fig.~\ref{fig:figS7}.
\label{app:V}

\section{Numerical Procedure}
\label{app:numerics}

We numerically compute the first-detection probabilities directly from the finite-size spectrum and the quantum renewal equation. For simplicity, we consider an even number $N$ of lattice sites and use the translational invariance of the Hamiltonian. Since the hopping matrix is circulant, its eigenvalues can be efficiently evaluated using a fast Fourier transform rather than by explicitly diagonalizing the $N\times N$ Hamiltonian. For an even periodic chain, the hopping vector is constructed as
\begin{align}
    g_d &=
    \begin{cases}
        d^{-\alpha}, & 1\leq d < N/2,\\
        (N/2)^{-\alpha}, & d=N/2,\\
        (N-d)^{-\alpha}, & N/2<d<N,
    \end{cases}
\end{align}
with $g_0=0$ indicating no self-hopping. The discrete Fourier transform of $g_d$ then gives the eigenenergies
\begin{align}
    E_{k_l}^{(N)} &= -\frac{\gamma}{\mathcal N_{\alpha,N}}
    \sum_{d=0}^{N-1}g_d \,e^{-ik_l d}  = -\frac{2\gamma}{\mathcal N_{\alpha,N}} \sum_{d=1}^{N/2-1}\frac{\cos(k_l d)}{d^\alpha}
    -  \frac{\gamma}{\mathcal N_{\alpha,N}}  \frac{\cos(k_lN/2)}{(N/2)^\alpha},
    \label{eq:numerical_spectrum}
\end{align}
where $  k_l=2\pi l/N$ with $l = 0,\ldots,N-1$ and  $ \mathcal N_{\alpha,N} = \sum_{d=1}^{N/2}d^{-\alpha} $ is the finite-size Kac normalization. The use of the fast Fourier transform reduces the cost of obtaining the complete spectrum from a direct diagonalization cost to $O(N\log N)$. For a localized initial state $\ket{0}$, all momentum eigenstates have equal weight $1/N$. The finite-size Loschmidt amplitude at the $n$-th measurement time is therefore evaluated from the spectrum~\eqref{eq:numerical_spectrum} as
\begin{align}
    \mathcal L_n^{(N)}  = \frac{1}{N}  \sum_{l=0}^{N-1} \exp\left(-iE_{k_l}^{(N)}n\tau\right).
\end{align}
Having obtained $\mathcal L_n^{(N)}$, we determine the first-detection amplitudes recursively from the quantum renewal equation (Eq.~(4) of the main text),
\begin{align}
    \phi_1 &= \mathcal L_1^{(N)},\\
    \phi_n
    &=
    \mathcal L_n^{(N)}
    -
    \sum_{j=1}^{n-1}
    \mathcal L_j^{(N)}\phi_{n-j},
    \qquad n\geq2.
\end{align}
The first-detection probability at the $n$-th attempt is computed as $F_n=|\phi_n|^2$ and the cumulative detection probability after $K$ measurements is calculated as $P_K=\sum_{n=1}^{K}F_n$.

\end{document}